*Under consideration for publication in J. Fluid Mech.*

# Nonlinear responses of the premixed V-flame subjected to dual-frequency disturbances

**Xiaozhen Jiang[1], Xiaokang Liu[2], Aimee S. Morgans[3], Jiaqi Nan[1], Lei Li[4], Tengyu Liu[5], Lijun Yang[1] and Jingxuan Li[1]**

[1]School of Astronautics, Beihang University, Beijing 100191, China

[2]Beijing Institute of Control Engineering, Beijing 100190, China.

[3]Department of Mechanical Engineering, Imperial College London, London, SW7 2AZ, UK.

[4]National Key Laboratory of Science and Technology on Aero-Engine Aero-thermodynamics, Research Institute of Aero-Engine, Beihang University, Beijing 102206, China.

[5]Hangzhou International Innovation Institute, Beihang University, Hangzhou 311115, China.

**Corresponding author:** J. Li, jingxuanli@buaa.edu.cn

The two-way interaction between the unsteady flame heat release rate (HRR) and acoustic waves can lead to combustion instability within combustors. Previous studies have typically characterised premixed flame responses to pure harmonic forcing, assuming dynamically linear or weakly nonlinear behaviour, to quantify flame-acoustic interactions. By combining third-order asymptotic analysis with numerical simulations of the $G$-equation, this study investigates the nonlinear response of laminar premixed V-flames subjected to dual-frequency velocity perturbations ($St_1$ and $St_2$, dimensionless frequencies). The positive correlation between disturbance propagation speed $u_c$ and frequency $St$ is captured by integrating a velocity-potential model with calibration against existing experimental data. The mechanism by which the disturbance at one forcing frequency, say $St_2$, affects the flame dynamic response at the other forcing frequency, $St_1$, is studied in detail. The perturbation at $St_2$ couples with that at $St_1$ to induce third-order nonlinear terms, giving rise to a non-monotonic suppression mechanism that smooths out the flame's spatial wrinkling owing to the positive correlation between $u_c$ and $St$. As a result, excitation at $St_2$ modifies the HRR response at $St_1$, delineating an effective region bounded on the left by the frequency threshold of the linear response and on the right by the aforementioned non-monotonicity. Within this region, excitation at $St_2$ can markedly attenuate the HRR gain at $St_1$ compared with the case where the flame is driven solely by the perturbation at $St_1$. For instance, once both perturbation amplitudes exceed a certain threshold, excitation at $St_2$ can attenuate the flame response at $St_1$ by more than 40% compared with the case without excitation at $St_2$. These findings contribute to the development of a quantitative framework for understanding how targeted frequency perturbations modulate the flame dynamics via nonlinear interactions, which may inform open-loop approaches for mitigating thermoacoustic instabilities in combustion chambers.

## 1. Introduction

With the development of lean combustion aero-engines and land-based gas turbines, combustion instability has proven to be a persistent problem that can cause serious damage (Poinsot 2017; Bailly *et al.* 2025). Combustion instability, also known as thermoacoustic instability, arises due to the interaction between flame heat release rate (HRR) fluctuations, oncoming flow perturbations upstream of the flame, and acoustic oscillations in the combustion chamber (Candel 2002). Its understanding and prediction require an understanding of how the flame HRR fluctuations respond to upstream velocity perturbations (Schuller *et al.* 2003), often characterised by the ratio of the normalised HRR fluctuation to the normalised velocity perturbation. At low perturbation levels, the flame typically responds linearly to flow perturbations, and this relation is quantified using a flame transfer function (FTF) (Fleifil *et al.* 1996). As the perturbation amplitudes increase, the flame response becomes nonlinear, and in the case of weak nonlinearity, is often characterised by a flame describing function (FDF) (Dowling 1997).

The FDF can be characterised through experiments (Noiray *et al.* 2008; Durox *et al.* 2009; Tian *et al.* 2024*a*) and numerical simulations (Qiao & Lv 2023; Krediet *et al.* 2013). A simpler alternative, which is able to quantitatively describe the FTF/FDF for laminar flames (Schuller *et al.* 2003; Orchini & Juniper 2016*b*; Tian *et al.* 2023; Qiao & Lv 2024; Tian *et al.* 2024*b*) and weakly turbulent flames (Palies *et al.* 2011; Jiang *et al.* 2022), derives the flame model from the $G$-equation proposed by Markstein (1964) for tracking infinitely thin flame fronts. Based on the linearisation of the $G$-equation for inclined flames, Schuller *et al.* (2003) derived an analytical unified model which included the convective effects of flow modulation upstream of the flame to quantify the FTF of laminar premixed flames. The obtained analytical solution was in good agreement with the experimental results (Schuller *et al.* 2002). Lieuwen (2005) developed a nonlinear framework for the $G$-equation to analyse the nonlinear dynamics of a premixed flame in response to harmonic velocity perturbations and was able to predict the experimentally observed nonlinear behaviour of the flame (Bellows *et al.* 2007).

A more comprehensive characterisation of the nonlinear flame response is possible by considering perturbations at two or even more frequencies. As well as offering a more complete understanding of the flame nonlinearity, such studies are also directly relevant to experiments (Balachandran *et al.* 2008; Lamraoui *et al.* 2011; Albayrak *et al.* 2018; Fang *et al.* 2021) and numerical simulations (Haeringer *et al.* 2019; Haeringer & Polifke 2019; Tathawadekar *et al.* 2021) exhibiting dual- or multiple-frequency oscillations.

Balachandran *et al.* (2008) conducted experiments investigating the nonlinear response of premixed flames at two different frequencies. With the introduction of subharmonics or higher harmonics, the formation and shedding of vortices change significantly. They elaborated on the possibility of suppressing instabilities by introducing additional excitation at carefully chosen frequencies to the flame. Based on the FDF framework, Boudy *et al.* (2011) further investigated nonlinear combustion instabilities in a multipoint injection combustor. Their results demonstrated that interactions between coexisting oscillation modes under dual-frequency forcing substantially alter the flame response, giving rise to nonlinear behaviours such as mode switching and hysteresis that cannot be captured by linear models. Lamraoui *et al.* (2011) experimentally investigated combustion instability in a turbulent vortex burner with two non-harmonically related unstable modes at 180

Hz and 280 Hz. The flame dynamics and corresponding combustion instabilities change significantly due to the presence of an additional disturbance. Similar modal interaction phenomena have also been reported in Rijke burner experiments, where the coupling between low-frequency flame pulsations and thermoacoustic modes gives rise to self-sustained amplitude modulation and beating (Weng *et al.* 2016).

In this context, the analysis related to the nonlinear response under dual- or multi-frequency excitation can be broadly categorised into higher harmonics and non-resonant frequencies of the flame response. In the former case, Haeringer *et al.* (2019) proposed an extended FDF based on the experimental phenomena of Albayrak *et al.* (2018) as an efficient way to include higher harmonics of the flame response. More specifically, Han *et al.* (2016) numerically examined the nonlinear response of a lean premixed flame subjected to two strong perturbations at 160 Hz and 320 Hz. The introduction of the higher-frequency forcing markedly altered HRR fluctuations, leading to a 70% reduction in the flame response at the fundamental frequency compared with single-frequency excitation. In a practical combustion system (Lamraoui *et al.* 2011), however, perturbations frequently involve frequency pairs that are not harmonically related, leading to more complex nonlinear flame responses than those arising from correlated frequencies. As linear analysis shows a pattern with more than one positive growth rate, it is difficult to determine the presence and stability of steady-state oscillations.

Ćosić *et al.* (2012) investigated open-loop forcing at non-resonant frequencies from the point of view of the nonlinear response of a flame to the superposition of two approximately sinusoidal input signals. They found that the reduction of the fundamental harmonic gain at unstable frequencies by additional forcing at non-resonant frequencies is a possible indicator of the success of open-loop control. Moeck & Paschereit (2012) proposed conditions for the existence and stability of single- or multi-mode steady-state oscillations and applied this method to a thermoacoustic model with two linearly unstable modes. In the case of considering static nonlinearities, Gelb & Velde (1968) obtained a large number of results in the study of two-frequency describing functions, most of whose properties are still applicable to dynamic nonlinearities.

Orchini & Juniper (2016*a*) presented the computation of a non-static flame dual-input describing function based on the $G$-equation model of a laminar conical flame. To perform harmonic balance analysis, they neglected the harmonic response at higher frequencies and assumed that the HRR response is dominated by components at the two input frequencies. It was able to predict the onset of the Neimark-Sacker bifurcation and determine the frequency of oscillations around the limit cycle. However, as noted by Orchini & Juniper (2016*a*), this simplification becomes less appropriate when nonlinear coupling between multiple inputs generates additional low-frequency components, such as flame responses at $|\omega_1 \pm \omega_2|$, where $\omega_1$ and $\omega_2$ denote the input angular frequencies. Along similar lines, Zheng *et al.* (2024) experimentally investigated the nonlinear flame response under dual-frequency excitation and explicitly examined the generation of interaction frequencies. Using a third-order asymptotic expansion of the $G$-equation under the assumption of a fixed disturbance convection velocity, they demonstrated the emergence of a pronounced difference-frequency component (e.g. $|\omega_1 - \omega_2|$), highlighting the second-order nonlinear interactions responsible for interaction-frequency generation.

Recently, the thermoacoustic response of hydrogen-enriched premixed flames subjected to non-harmonically related dual-mode forcing has been investigated using an FDF coupled with a low-order network model (Tian *et al.* 2024*b*). The results demonstrated that dual-frequency coupling can substantially modify the nonlinear flame dynamics, leading to mode competition, suppression or amplification of the primary response and the emergence of difference-frequency oscillations. Building upon these findings, Tian *et al.* (2026) showed that such difference-frequency oscillations may become dominant through nonlinear modal interactions that are not captured by conventional single-input FDFs. To overcome this limitation, they introduced a dual-input FDF framework that enables frequency-domain modelling of cross-frequency coupling and difference-frequency generation, thereby improving predictive capability in multi-mode thermoacoustic systems.

These studies demonstrate that experimental, theoretical and numerical approaches involving dual- or multi-frequency perturbations provide deeper insights into flame nonlinearities and offer promising strategies for controlling combustion instabilities. However, despite the recognised influence of non-resonant frequency disturbances on the primary flame response, the underlying mechanisms governing the interaction between uncorrelated frequencies remain insufficiently understood. Specifically, in most of these studies, the flame module is treated as a "black box" within the thermoacoustic network, thereby overlooking the underlying mechanisms responsible for the nonlinear flame response under dual- or multi-frequency perturbations. This limits physical understanding of the nonlinear flame response and further restricts its application in more practical cases.

As mentioned above, there is a deviation in the nonlinear flame response to a perturbation at frequency $St_1$ ($St$ is the Strouhal number, defined by the angular frequency multiplied by the spatial distance divided by the mean bulk velocity), when there is also a simultaneous excitation at a different frequency $St_2$, compared with when $St_1$ is the only excitation frequency. In this study, the impact of the disturbance at $St_2$ on the nonlinear flame response at $St_1$ will be investigated. While Zheng *et al.* (2024) examined the generation of a flame response at different frequencies induced by excitation at $St_2$, the present study instead focuses on the suppression of the flame response at the primary frequency. This analysis will be conducted for general cases where the frequency, amplitude and phase of the perturbation at $St_2$ are independent of those at $St_1$. In general, the nonlinear response of a V-flame to excitation is much more pronounced than that of a conical flame as the perturbation amplitude varies, as illustrated by the experiments of Durox *et al.* (2009). It is more appropriate to explore the above targets within the framework of V-flames with a more intense nonlinear response, so that it can be more clearly identified how the nonlinear response of the flame to excitation changes with the introduction of additional excitation at different frequencies.

In this paper, the spatial distribution of flame kinematics and the HRR response will be considered to quantify and account for the full nonlinearity induced by the dual-frequency excitation. The nonlinear results for the flame response are derived by considering a low-order asymptotic analysis (up to the third order in normalised excitation amplitude) applied to numerical computation of the flame based on the model framework of the $G$-equation. Key empirical parameters are obtained from the experimental data and combined with results from a velocity potential model to derive the dependence of the propagation speed of the disturbance on $St$. These are described in § 2. The mechanisms by which the perturbation at $St_2$ affects the spatial kinematics of the flame at $St_1$ are discussed. These results are presented in § 3. The role of the perturbation at $St_2$ in the global flame response at $St_1$ is quantified, and the corresponding mechanisms are carefully discussed in § 4.

## 2. Formulation

This section presents analytical and numerical methods for determining the nonlinear acoustic response of laminar premixed V-flames to dual-frequency perturbations. As shown in figure 1, using the cylindrical coordinate system $x-r-\theta$ ($x$, $r$ and $\theta$ correspond to the axial, radial and angular directions, respectively), an axisymmetric V-flame is anchored on the rim of the burner and subjected to the mean bulk velocity $\bar{u}$ and two velocity perturbations $u_1'$ and $u_2'$ with different frequencies and amplitudes. The instantaneous positions of the perturbed premixed flame can be captured by the $G$-equation introduced by Markstein (1964) to track the propagation of unsteady convected interfaces:

$$\frac{\partial G}{\partial t} + \boldsymbol{u} \cdot \nabla G = S_L \left| \nabla G \right| \tag{2.1}$$

where $G$ is a scalar level-set function, $t$ is time, $\boldsymbol{u}$ is the velocity vector of a given flow field, i.e., $\boldsymbol{u} = (u, u_r, u_\theta)$ and $S_L$ is the laminar flame speed. The $G$-equation (2.1) in this work assumes that: (i) an infinitely thin flame front ($G = 0$) separates the unburned reactant region ($G < 0$) and burned gas region ($G > 0$); (ii) the flame laminar displacement speed is constant, which assumes a spatially and temporally constant equivalence ratio and omits the flame curvature effect; (iii) the flow field upstream of the flame front can be predefined, and both heat diffusion and thermal expansion are neglected, implying weak density change across the flame front; and (iv) the flame height is a single-valued function of the radial coordinate.

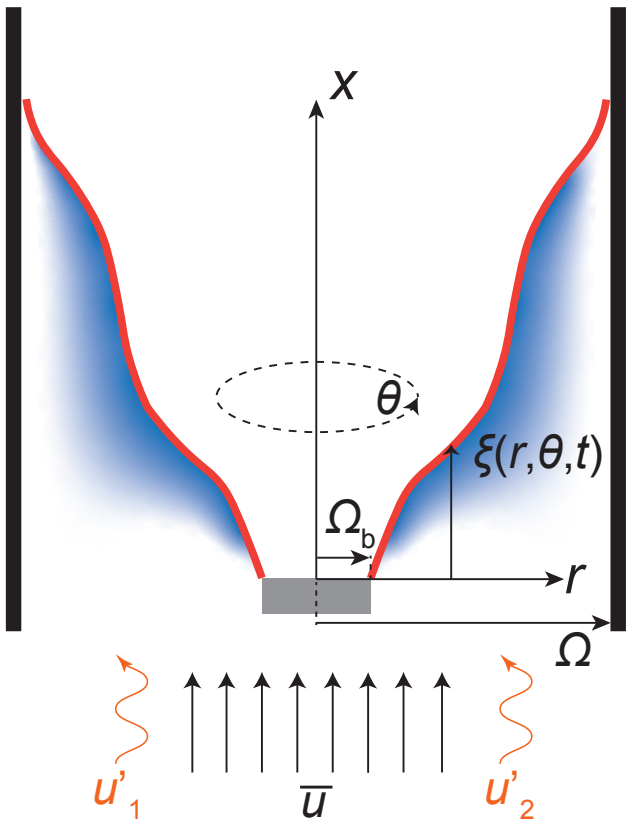

Figure 1. Sketch of the perturbed V-flame subjected to dual-frequency disturbances. Here, $x$, $r$, and $\theta$ denote the spatial coordinates; $\xi(r,\theta,t)$ is the flame-front location; $u_1'$ and $u_2'$ are the perturbations at $St_1$ and $St_2$, respectively; $\bar{u}$ is the mean bulk velocity; $\Omega$ is the outer burner radius; and $\Omega_b$ is the inner burner radius.

Equation (2.1) is rewritten as:

$$\frac{\partial G}{\partial t} + u\frac{\partial G}{\partial x} + u_r\frac{\partial G}{\partial r} + \frac{u_\theta}{r}\frac{\partial G}{\partial \theta} = S_L\sqrt{\left(\frac{\partial G}{\partial x}\right)^2 + \left(\frac{\partial G}{\partial r}\right)^2 + \left(\frac{1}{r}\frac{\partial G}{\partial \theta}\right)^2} \tag{2.2}$$

According to assumption (iv), $G\,(x, r, \theta, t)$ can be transformed into explicit form with respect to the instantaneous flame front $\xi\,(r, \theta, t)$. This means that, for a given radial position, only one axial position of the flame front is possible at a given instant. Although this is not entirely consistent with experimental observations, which show large sharp angles and different axial positions of the flame front for a given radius, it captures the main features of nonlinear flame response to excitations in both laminar and weakly turbulent cases (Hemchandra *et al.* 2011; Jiang *et al.* 2022):

$$G\,(x, r, \theta, t) = x - \xi\,(r, \theta, t) \tag{2.3}$$

Substituting equation (2.3) into equation (2.2), the flame front is $\theta$-independent due to the angular symmetry of the mean flow and incoming flow perturbations (Lieuwen 2005; Preetham *et al.* 2008). After further simplification, lower-order models based on the $G$- equation have been proposed in the literature (Schuller *et al.* 2003; Preetham *et al.* 2008). Assuming that the incoming mean bulk velocity has only an axial component and lacks a corresponding radial component ($\bar{u}_r = 0$), the governing equation for the flame front location $\xi$ is then recast as:

$$\frac{\partial \xi}{\partial t} - u + u_r' \frac{\partial \xi}{\partial r} = -S_L \sqrt{1 + \left(\frac{\partial \xi}{\partial r}\right)^2} \tag{2.4}$$

where $u_r'$ is the radial velocity perturbation.

With the flame anchored on the burner rim, a boundary condition for equation (2.4) can be expressed as:

$$\xi \left(r = \Omega_b, t\right) = 0 \tag{2.5}$$

where $\Omega_b$ is the radius of the centre body. In the remainder of this paper, all parameters are normalised, that is, velocities are normalised by the mean bulk velocity $\bar{u}$, spatial coordinates are normalised by the burner radius $\Omega$ and time by $\Omega/\bar{u}$.

### 2.1. *Frequency-dependent model for incoming velocity perturbations*

For the upstream velocity field of the flame, once the axial velocity perturbation $u'$ is specified, the radial perturbation $u_r'$ is not an independent quantity but is determined from the continuity equation under the incompressible and irrotational flow assumptions. This results in a radial velocity component that varies linearly with $r$ and depends on forcing frequency, as discussed in Cuquel *et al.* (2011). Consequently, inclusion of the radial perturbation term in equation (2.4) leads to non-physical results for the V-flame response. Specifically, the intensity of the $u_r'$ at the tip of the V-flame is excessively high and further amplifies the flame response within the context of the $G$-equation, especially at high frequencies, which is physically unrealistic. Therefore, zero radial fluctuation is assumed ($u_r' = 0$) in equation (2.4) despite the discontinuity in the prescribed velocity field. It should be noted that this treatment may represent a limitation of the present framework because the configuration considered in Cuquel *et al.* (2011) differs from the present flame configuration, where heat release and finite flame extent may affect the radial convection dynamics. The issue related to neglecting $u_r'$ is not encountered in the case of the conical flame that is axially perturbed because of its centred flame tip. The influence of the neglected radial perturbation term $u_r'$ will be reconsidered using an alternative approach in the subsequent model construction.

Therefore, the velocity can be decomposed into a mean component and two perturbations of distinct frequencies and amplitudes originating from the thermoacoustic system:

$$u\left(St_1, St_2, x, t\right)/\bar{u} = 1 + \epsilon_1 C_1\left(St_1, x, t\right) + \epsilon_2 C_2\left(St_2, x, t, \delta\right) \tag{2.6}$$

Herein, $C$ is the individual component of acoustic excitation and $\epsilon$ denotes the non-dimensional amplitude; $St = \omega\Omega/\bar{u}$ is the normalised angular frequency; $\omega$ is the angular frequency. The subscripts “1” and “2” denote the perturbations at $St_1$ and $St_2$, respectively, and $\delta$ denotes the initial phase difference between two perturbing components. The specific forms of the perturbation sources are given by

$$\begin{cases} C_1\left(St_1, x, t\right) = \cos\left[St_1\left(K_1 x - t\right)\right] \\ C_2\left(St_2, x, t, \delta\right) = \cos\left[St_2\left(K_2 x - t\right) + \delta\right] \end{cases} \tag{2.7}$$

Here, $K_1$ and $K_2$ characterise the disturbance convection velocity $u_c$ and are expressed as $\bar{u}/u_c$. It is frequently simplistically assumed that perturbations convect either with the mean volumetric velocity or with the sound speed, corresponding to $K = 1$ or $K \to 0$, respectively. Although this assumption is relatively straightforward, it typically fails to accurately reflect the actual convective behaviour observed in experiments or numerical simulations.

For instance, experimental studies (Birbaud *et al.* 2006; Karimi *et al.* 2009; Yang *et al.* 2021) and numerical simulations (Blanchard *et al.* 2015) have highlighted that the propagation of disturbances changes significantly as the modulation frequency upstream of the flame changes. Recent work (Steinbacher & Polifke 2022) has emphasised that the need to calibrate $u_c$ arises from neglecting thermal diffusion and thermal expansion in the $G$-equation model. The propagation of the disturbance along the flame sheet and the feedback to the velocity disturbances are responsible for $u_c$. Therefore, $K$ in the $G$-equation should be considered as an input parameter that requires calibration. However, in reality, propagation velocities cannot be an independent "input" because they are determined by the closed-loop feedback of flame-flow interactions. This study proposes a simplified derivation (referred to as low-order modelling), which builds upon and extends the work of Birbaud *et al.* (2006). Its objective is to avoid the aforementioned conflicts while encapsulating the key characteristics of the relationship between $St$ and $K$.

Within the velocity-potential framework, the radial velocity perturbation $u_r'$ enters the formulation implicitly through its contribution to the $St$-$K$ relation, thereby influencing the flame kinematics without appearing explicitly in the flame-front equation. With $z = 1 - r$, a coordinate transformation is used to simplify the derivation process. Assuming the flow is incompressible and irrotational upstream of the flame, the velocity potential $\psi$ satisfies the Laplace equation and is written in cylindrical coordinates as

$$\frac{\partial^2 \psi}{\partial x^2} + \frac{1}{z}\frac{\partial \psi}{\partial z} + \frac{\partial^2 \psi}{\partial z^2} = 0. \tag{2.8}$$

Due to the presence of disturbances, $\psi$ is assumed to be in a wave-like form (i.e., $K_x$, corresponding to axial perturbation propagation, is assumed to be unity):

$$\psi = g(z)\exp\left[\mathrm{i}St(x - t)\right]. \tag{2.9}$$

Substituting equation (2.9) into equation (2.8), the ordinary differential equation for the radial velocity potential distribution becomes

$$\frac{\mathrm{d}^2 g(z)}{\mathrm{d}z^2} + \frac{1}{z}\frac{\mathrm{d}g(z)}{\mathrm{d}z} - St^2 g(z) = 0. \tag{2.10}$$

The general solution is

$$g(z) = A_1 I_0(Stz) + A_2 K_0(Stz). \tag{2.11}$$

where $I_0$ and $K_0$ are the zeroth-order modified Bessel functions.

The radial velocity component vanishes on the centreline $z = 0$, i.e., $\partial\psi/\partial z|_{z=0} = 0$, resulting in $A_2 = 0$.

Following the approach of Birbaud *et al.* (2006), it is assumed that the flame front is nearly vertical, i.e., the flame half-angle is small. Under this assumption, flame wrinkles are considered to affect the radial velocity. One may write

$$\left.\frac{\partial \psi}{\partial z}\right|_{z=\eta(x)} = B\cos\alpha\exp\left[\mathrm{i}St(x - t)\right]. \tag{2.12}$$

where $\eta(x)$ is the inverse function of $\xi(z)$, denoting the flame front; $B$ denotes a pre-exponential coefficient and $\alpha$ is the half-angle of the flame tip.

Substituting equation (2.9) into equation (2.12), one obtains

$$g(z) = BI_0(Stz)/I_1(St\eta(x)), \tag{2.13}$$

corresponding to the velocity potential solution as follows:

$$\psi(St, z, x) = \frac{BI_0(Stz)}{I_1[St\eta(x)]} \exp[\mathrm{i}St(x-t)]. \tag{2.14}$$

Thus, the relative velocity potential $\vartheta$, defined as the ratio of the velocity potential on the reactant side to that on the flame front, is given by

$$\vartheta(St, z, x) = \frac{I_0(Stz)}{I_0[St\eta(x)]}. \tag{2.15}$$

It is shown in kinematic analyses and experiments that the acoustic wave can produce convective wrinkling on the flame front subjected to acoustic modulation, i.e., $K_\eta = 1$. Therefore, the ratio of localised perturbation propagation velocities ($\kappa = \bar{u}/u_c(z,x)$) at different spatial locations upstream of the flame can be derived using the parameter $\vartheta$.

To further relate $\kappa$ to the relative velocity potential, an important empirical value of $\Lambda$ is used to relate the model results to the experimental data. If $\vartheta$ is smaller than the threshold $\Lambda$, the perturbation speed is assumed to be the speed of sound, i.e.,

$$\kappa(St, z, x) = 0 \text{ when } \vartheta(St, z, x) < \Lambda. \tag{2.16}$$

Conversely, when $\vartheta$ exceeds $\Lambda$, the perturbation is convected by the mean bulk velocity, i.e.,

$$\kappa(St, z, x) = 1 \text{ when } \vartheta(St, z, x) \geq \Lambda. \tag{2.17}$$

Then the essential issue is to determine the threshold value $\Lambda$ from the experimental results. In the framework of conical flames, Birbaud *et al.* (2006) assumed a threshold of $\Lambda = 0.1$, and the corresponding modelling results were shown to be in good agreement with experimental velocity perturbations measured upstream of the flame by particle image velocimetry (PIV). Recently, Yang *et al.* (2021) conducted similar experiments and found that $\Lambda = 0.2$ represented the experimental results better. Figure 2 presents a comparison of the modelling results for two different $\Lambda$ cases with the experiments, where the modelling of the conical flame is also derived from equations (2.8) ~ (2.17), but here $z$ is $r$ instead of $1-r$. For details on obtaining experimental data from Birbaud *et al.* (2006), see Appendix A. The physical meaning of $\Lambda$ can be interpreted as follows: $\Lambda$ strongly influences the propagation behaviour of these perturbations by altering $\kappa$ through equations (2.16) and (2.17). As $\Lambda$ increases, $u_c$ approaches the speed of sound more readily. Based on comparison with the experimental data, $\Lambda = 0.2$ provides better agreement and is therefore adopted for the remainder of this study.

It should be noted that the existing studies (Birbaud *et al.* 2006; Yang *et al.* 2021) pertain exclusively to conical flames, and no comparable results are currently available for the determination of $\Lambda$ in V-flames. Therefore, in the case of V-flames, $\Lambda$ must be inferred from experimental data for conical flames in order to more realistically reveal how the upstream value of $\kappa$ varies with spatial position. The V-flame considered here is a simplified version of an inverted conical laminar flame, with the effect of vorticity on flame kinematics being neglected. This greatly reduces the differences between V-flames and conical flames, thereby justifying the application of the $\Lambda$ value derived from conical flames to V-flames. Indeed, these simplifications appear to be reasonable, as will be demonstrated through subsequent experiments and model comparisons (see § 4).

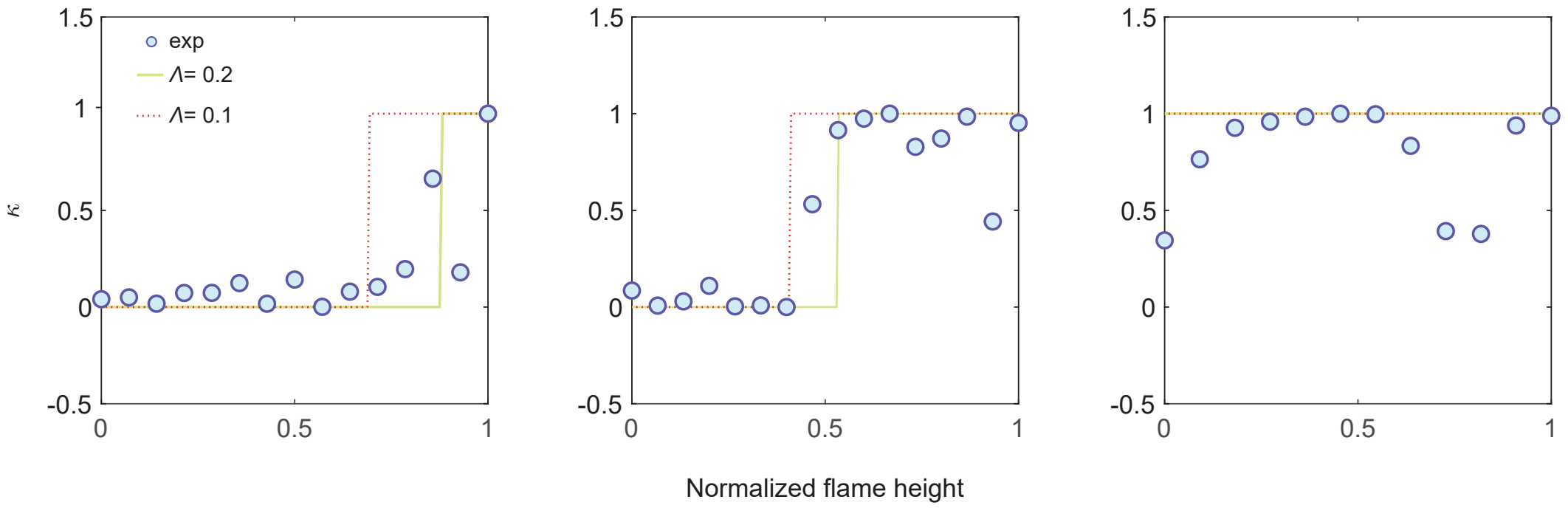

Figure 2. Comparison between the model predictions and experimental results for the ratio of the localised perturbation propagation velocity $\kappa$ along the centreline of a conical flame. From left to right: $f$ = 250 Hz, $f$ = 130 Hz and $f$ = 50 Hz. Experimental data reproduced from Birbaud *et al.* (2006).

A global parameter $K$, which characterises the overall properties of the reactant side, is defined as:

$$K\left(St\right)=\int_{\theta=0}^{2\pi}\mathrm{d}\theta\int_{x}\int_{z}\kappa\left(St,z,x\right)z\mathrm{d}x\mathrm{d}z\bigg/\int_{\theta=0}^{2\pi}\mathrm{d}\theta\int_{x}\int_{z}z\mathrm{d}x\mathrm{d}z. \tag{2.18}$$

The relation between $St$ and $K$ is illustrated in figure 3. At low frequencies, inflow disturbances propagate with the mean bulk velocity. As the modulation frequency increases, the disturbance propagation speed also increases. When the forcing frequency becomes sufficiently high, $u_c$ approaches the speed of sound. At intermediate frequencies, perturbation propagation exhibits mixed characteristics; $u_c$ varies with spatial position and is no longer uniform. The corresponding spatially averaged value, $\bar{u_c}$, lies between $\bar{u}$ and $c$. These three distinct types of frequency-dependent behaviour are qualitatively consistent with experimental observations (Birbaud *et al.* 2006) and numerical simulations (Blanchard *et al.* 2015). It is important to note that this method cannot capture cases where $u_c < \bar{u}$ ($K > 1$), as observed in certain spatial regions at specific frequencies in different cases (Birbaud *et al.* 2006; Karimi *et al.* 2009; Blanchard *et al.* 2015). However, most analytical studies on laminar premixed flames typically assume $K = 0$ or $K = 1$, ignoring its dependence on modulation frequency (Zheng *et al.* 2024). The present approach improves upon previous models by quantitatively linking $K$ to the influence of $St$ within the range from 0 to 1.

## 2.2. *Low-order asymptotic analysis of the flame front*

An asymptotic analysis of the governing equation is performed to reveal the underlying mechanisms of the nonlinear interaction between the two perturbations. Asymptotic analysis is extended up to third order in the excitation amplitudes $\epsilon_1$ and $\epsilon_2$, as follows:

$$\begin{aligned}\xi(r,t) = \; & \underbrace{\xi_0(r)}_{\text{Steady term}} + \underbrace{\epsilon_1\xi_1(r,t) + \epsilon_2\xi_2(r,t)}_{\text{Linear terms}} \\ & \underbrace{+\epsilon_1{}^2\xi_{1,1}(r,t) + \epsilon_2{}^2\xi_{2,2}(r,t) + \epsilon_1{}^3\xi_{1,1,1}(r,t) + \epsilon_2{}^3\xi_{2,2,2}(r,t)}_{\text{Self-nonlinear terms}} \\ & \underbrace{+\epsilon_1\epsilon_2\xi_{1,2}(r,t) + \epsilon_1{}^2\epsilon_2\xi_{1,1,2}(r,t) + \epsilon_1\epsilon_2{}^2\xi_{1,2,2}(r,t)}_{\text{Mutual-nonlinear terms}} \\ & + O\left[(\epsilon_1,\epsilon_2)^4\right]\end{aligned} \tag{2.19}$$

where the subscript “0” denotes the steady flame configuration. The subscripts “1” and “2” identify response components associated with excitations at frequencies $St_1$ and $St_2$, respectively. For the flame front $\xi$, the number of subscript indices indicates the order of the flame response in the perturbation expansion. A single index (e.g., $\xi_1$ or $\xi_2$) represents a first-order flame response, two indices denote second-order contributions, and three indices correspond to third-order terms. The specific combination of indices identifies the frequency origin of the nonlinear flame response. For instance, $\xi_{1,1}$ represents a second-order self-nonlinear flame response arising from excitation at $St_1$, whereas $\xi_{1,2}$ denotes a second-order mutual-nonlinear flame response resulting from the interaction between response components associated with excitations at $St_1$ and $St_2$. Similarly, $\xi_{2,2,2}$ corresponds to a third-order self-nonlinear flame response associated exclusively with excitation at $St_2$, while $\xi_{1,2,2}$ represents a third-order term resulting from the interaction between a first-order response to excitation at $St_1$ and a second-order response to excitation at $St_2$. Each term is proportional to the corresponding product of perturbation amplitudes (e.g., $\epsilon_1^2$, $\epsilon_1\epsilon_2$, or $\epsilon_1\epsilon_2^2$), consistent with the associated subscript structure.

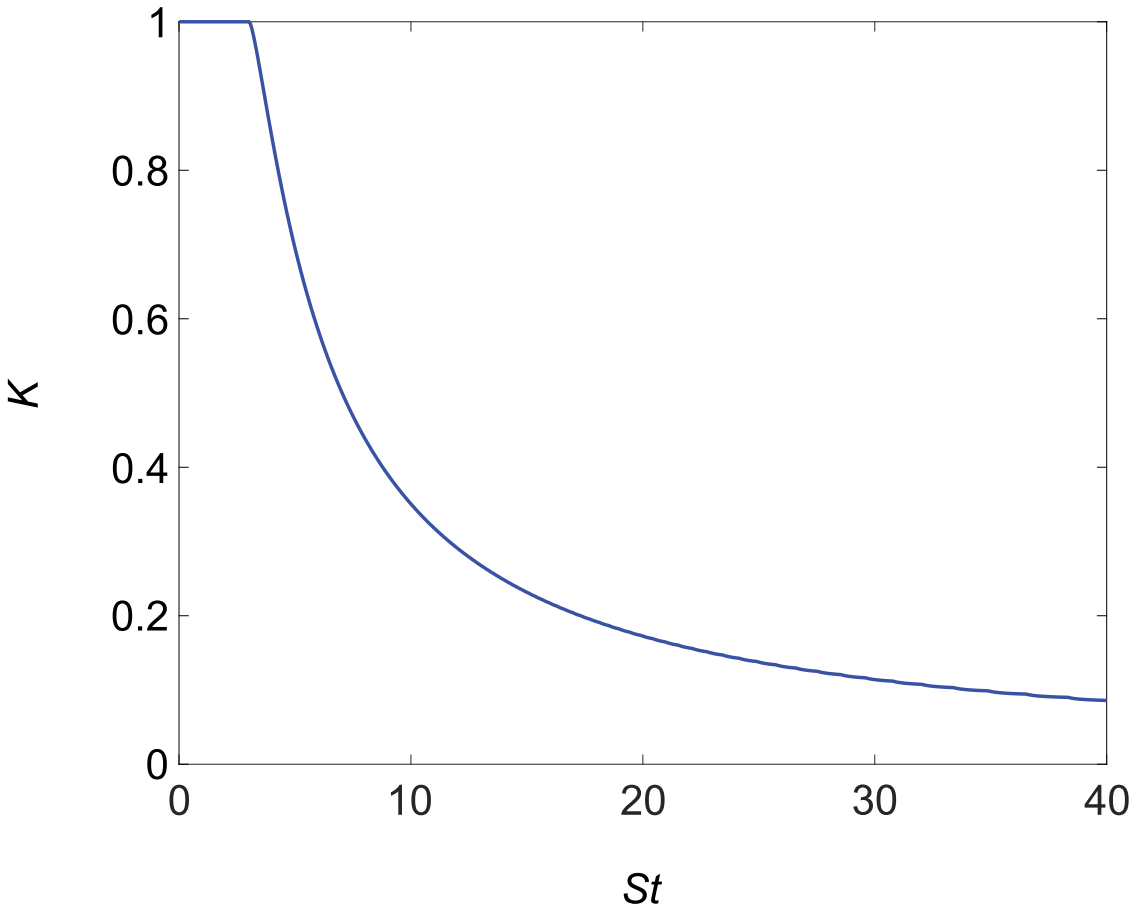

Figure 3. Dependence of $K$ on $St$ for the V-flame. Here, $\Lambda = 0.2$, $\Omega_b = 0$.

These ten terms can therefore be grouped into three categories: linear, self-nonlinear and mutual-nonlinear contributions. The linear and self-nonlinear terms correspond to responses induced by individual excitations at $St_1$ or $St_2$, whereas the mutual-nonlinear terms arise from the coupled action of both excitations.

The steady-term solution for the V-flame can be easily obtained as follows:

$$\xi_0(r) = r\cot\alpha. \tag{2.20}$$

By substituting equation (2.19) into the nonlinear source term on the right-hand side (RHS) of equation (2.4) and applying a Taylor expansion up to third order, the nonlinear term can be expressed as:

$$\sqrt{1+\left(\frac{\partial\xi}{\partial r}\right)^2} \approx \iota_0 + \iota_1\left(\frac{\partial\xi}{\partial r}-\frac{\partial\xi_0}{\partial r}\right) + \frac{1}{2}\iota_2\left(\frac{\partial\xi}{\partial r}-\frac{\partial\xi_0}{\partial r}\right)^2 \tag{2.21}$$

where

$$\begin{aligned}
\iota_0 &= \left[1+\left(\frac{\partial\xi_0}{\partial r}\right)^2\right]^{1/2};\\
\iota_1 &= \frac{\partial\xi_0}{\partial r}\left[1+\left(\frac{\partial\xi_0}{\partial r}\right)^2\right]^{-1/2};\\
\iota_2 &= \left[1+\left(\frac{\partial\xi_0}{\partial r}\right)^2\right]^{-1/2} - \left(\frac{\partial\xi_0}{\partial r}\right)^2\left[1+\left(\frac{\partial\xi_0}{\partial r}\right)^2\right]^{-3/2}.
\end{aligned} \tag{2.22}$$

The nonlinear source term has a clear physical interpretation, representing the propagation of the flame along its normal direction.

Following the standard procedure of asymptotic analysis, the detailed solution of each term in equation (2.19) can be obtained. The specific forms of these terms are provided in Appendix B. Focusing on the main issue addressed in this work, the terms were analysed concisely, three of which (equations (B 12), (B 16) and (B 20)) were found to be directly related to the flame kinematics under the perturbation at $St_1$. The specific expression is given by:

$$\begin{aligned}
\xi_{St_1}(r,t) &= \underbrace{\epsilon_1\xi_1(r,t)}_{\text{Linear solution}} + \underbrace{{\epsilon_1}^3\xi_{\mathrm{I},1,1,1}(r,t)}_{\text{Self-nonlinear solution}} + \underbrace{\epsilon_1{\epsilon_2}^2\xi_{\mathrm{I},1,2,2}(r,t)}_{\text{Mutual-nonlinear solution}}\\
&= \epsilon_1\mu_1(r)\cos\left[St_1 t+\chi_1(r)\right] + {\epsilon_1}^3\mu_{\mathrm{I},1,1,1}(r)\cos\left[St_1 t+\chi_{\mathrm{I},1,1,1}(r)\right] +\\
&\quad \epsilon_1{\epsilon_2}^2\mu_{\mathrm{I},1,2,2}(r)\cos\left[St_1 t+\chi_{\mathrm{I},1,2,2}(r)\right]\\
&= \mu_{St_1}(r)\cos\left[St_1 t+\chi_{St_1}(r)\right]
\end{aligned} \tag{2.23}$$

where the subscript "$St_1$" denotes the solution at $St_1$, $\mu(r)$ and $\chi(r)$ denote the amplitude and phase of the flame spatial response.

These three components have clear physical interpretations: the first term represents the linear response of the flame kinematics to excitation at $St_1$, while the remaining two terms arise from the nonlinearity of flame kinematic restoration. The derivation of the two third-order terms in equation (2.23) can be determined based on low-order asymptotic analysis. The former (called the self-nonlinear term, $\xi_{\mathrm{I},1,1,1}$) is induced by perturbation at $St_1$, and the second term (called the mutual-nonlinear term, $\xi_{\mathrm{I},1,2,2}$) arises from simultaneous excitations at both $St_1$ and $St_2$. The term $\xi_{\mathrm{I},1,2,2}$ is significant, as it directly modifies the flame response behaviour at $St_1$ under perturbation at $St_2$. In other words, $\xi_{\mathrm{I},1,2,2}$ represents the variation of flame kinematics at $St_1$ when an excitation at frequency $St_2$ is applied. This demonstrates that asymptotic analysis provides an efficient approach to assess the influence of linear, self-nonlinear and mutual-nonlinear responses arising from perturbations at $St_2$ and $St_1$ on flame kinematics at $St_1$, ultimately enhancing the understanding of their role in flame front tracking.

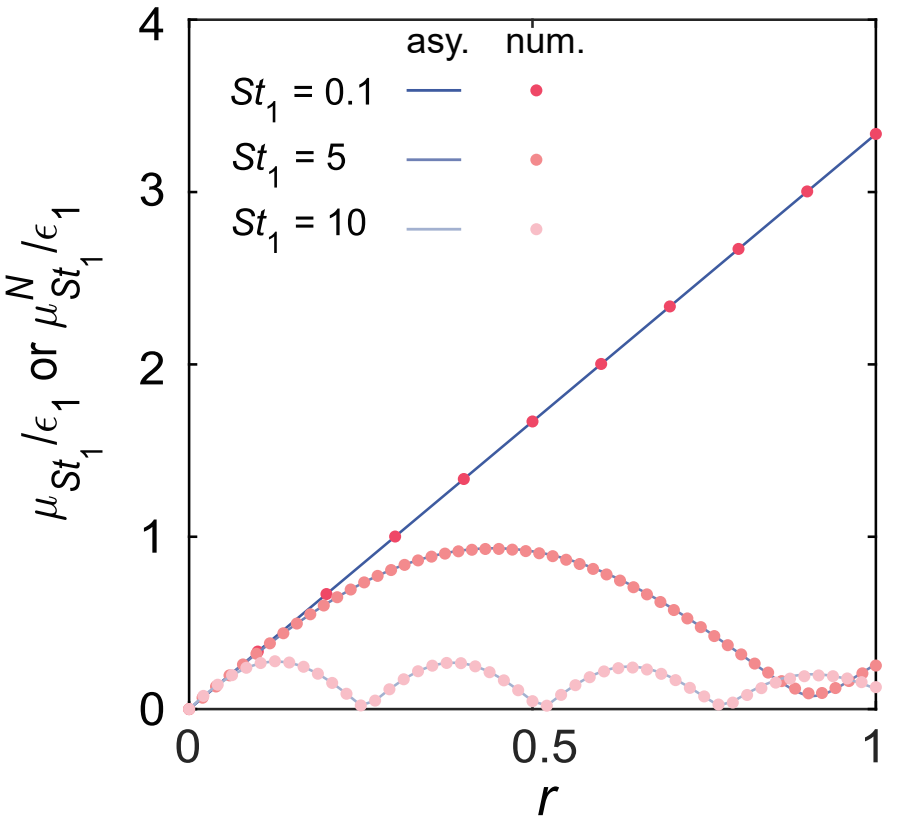

Figure 4. Comparison between asymptotic and numerical predictions of the flame-front response at $St_1$ for the V-flame. Solid lines and markers denote the asymptotic and numerical results, respectively. Here, $\epsilon_1 = 0.2$, $\epsilon_2 = 0.1$, $\delta = \pi/4$, $R = 1.6$, $R = St_2/St_1$, $\cot\alpha = 3$, and $\Omega_b = 0$. The superscript "$N$" denotes numerical results.

## 3. Flame kinematics

In this section, the spatial kinematics of V-flames at $St_1$ under two-frequency excitations are analysed asymptotically. Numerical methods are also employed by embedding the relation between $K$ and $St$ to directly solve for the flame front $\xi$, and the corresponding results are presented to verify the analytical findings. Spatial discretisation is performed using the seventh-order weighted essentially non-oscillatory (WENO7) scheme (Jiang & Peng 2000). Because nodes are defined only within the computational domain, the spatial derivatives at the boundary nodes are discretised using the fifth-order accurate upwind-differencing scheme. For time integration, a five-stage, fourth-order strong stability-preserving Runge-Kutta method (SSP-RK45) (Shu 2009) is used. The local Lax-Friedrichs (LLF) scheme (Jiang & Peng 2000) is applied to enhance stability. The spatial grid spacing is set to $2\pi St/(200K)$, and the temporal step size is determined according to the Courant-Friedrichs-Lewy (CFL) condition with CFL = 0.2.

It is assumed that the response includes all possible frequency interactions between $St_1$ and $St_2$, while the sub-harmonics of the input frequencies are neglected. The double Fourier series expansion (Orchini & Juniper 2016*a*) is employed to represent the flame front, which can be written as

$$\xi(r,t) = \sum_m \sum_n \mu^N_{mSt_1+nSt_2}(r)\cos\left[(mSt_1+nSt_2)t+\chi^N_{mSt_1+nSt_2}(r)\right]+\xi_0(r) \tag{3.1}$$

where $m, n \in \mathbb{Z}$, and the superscript "$N$" denotes numerically obtained results. As mentioned above, this study primarily focuses on the flame kinematics at $St_1$; hence, only the numerical results corresponding to $St_1$ are extracted from equation (3.1).

Figure 4 compares the relative magnitudes of the numerical and asymptotic solutions for the flame-front response at $St_1$ under dual-frequency excitation. The spatial response amplitudes of the flame at $St_1$ obtained by asymptotic analysis agree well with those determined numerically across different perturbation cases. This demonstrates that asymptotic analysis is reliable for capturing the flame kinematics at $St_1$ under dual-frequency excitation (through equation (2.23)) and for analysing their formation path, i.e., the perturbation at $St_2$ modifies the flame spatial response at $St_1$ solely through the $\xi_{I,1,2,2}$ term. In contrast, while the full numerical approach can produce reliable ensemble results at $St_1$, it cannot classify and individually analyse these results based on their physical significance.

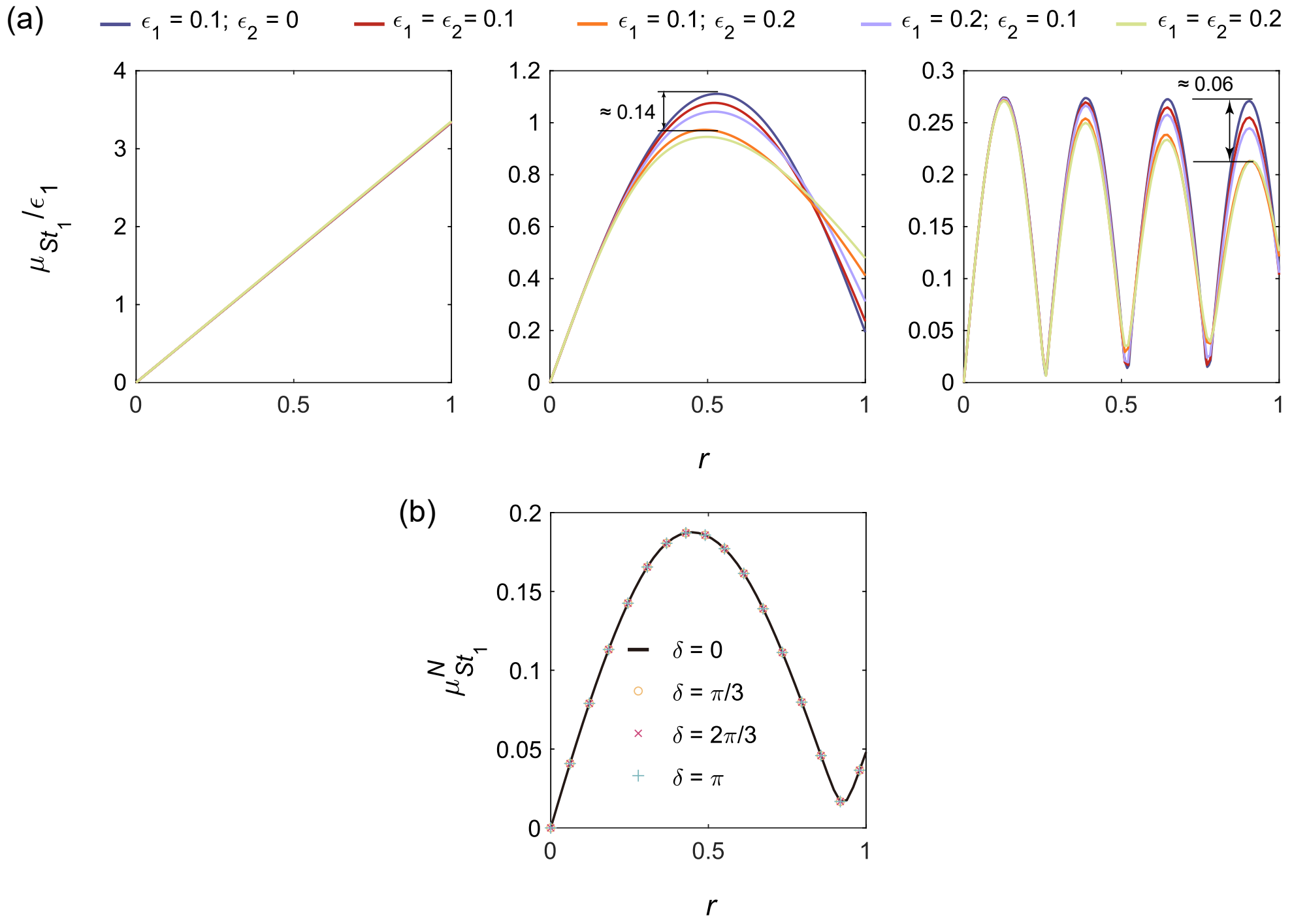

Figure 5. Relative flame-front response of the V-flame at $St_1$. (a) Effect of excitation amplitude for $\delta = 0$. From left to right: $St_1 = 0.1$ and $R = 0.3$; $St_1 = 4.7$ and $R = 0.9$; $St_1 = 10$ and $R = 0.6$. (b) Effect of the initial phase difference between the two excitations for $\epsilon_1 = \epsilon_2 = 0.1$, $St_1 = 5$, and $R = 1.2$. The superscript "$N$" denotes numerical results. Here, $\cot\alpha = 3$, $R = St_2/St_1$, and $\Omega_b = 0$.

When the forcing frequency is low or the flame position is located near the flame holder, the flame exhibits a purely linear response, and its characteristics are only marginally influenced by the perturbation frequency (see figure 4) and amplitude (see figure 5(a)). A previous study (Shanbhogue *et al.* 2009) has reported similar results when the modulation frequency is low or the spatial position is close to the flame holder. The characteristics of the flame response are all governed by a spatial interference wavelength defined as $1/[2\pi St(1-K)]$. The reduced length scale may be either larger or smaller than the interference wavelength. In the former case, the spatial response amplitude of the flame exhibits multiple peaks, as shown in the case of $St_1 = 10$ in figure 5(a). With decreasing $St_1$, the interference wavelength increases, causing the multiple peaks to disappear and giving rise to a single peak or no peak, as observed for $St_1 = 4.7$ and $St_1 = 0.1$, respectively, in figure 5(a). These characteristics were also observed in certain experiments (Shanbhogue *et al.* 2009). As the amplitude of the dual-frequency excitation varies, the perturbation at $St_2$ exerts a stronger suppressive effect on the flame kinematic response at $St_1$ than the excitation at $St_1$. In addition, the degree of suppression of the spatial response induced by the perturbation varies non-monotonically with excitation frequency (first increasing and then decreasing), indicating that the suppression is most pronounced at a critical frequency.

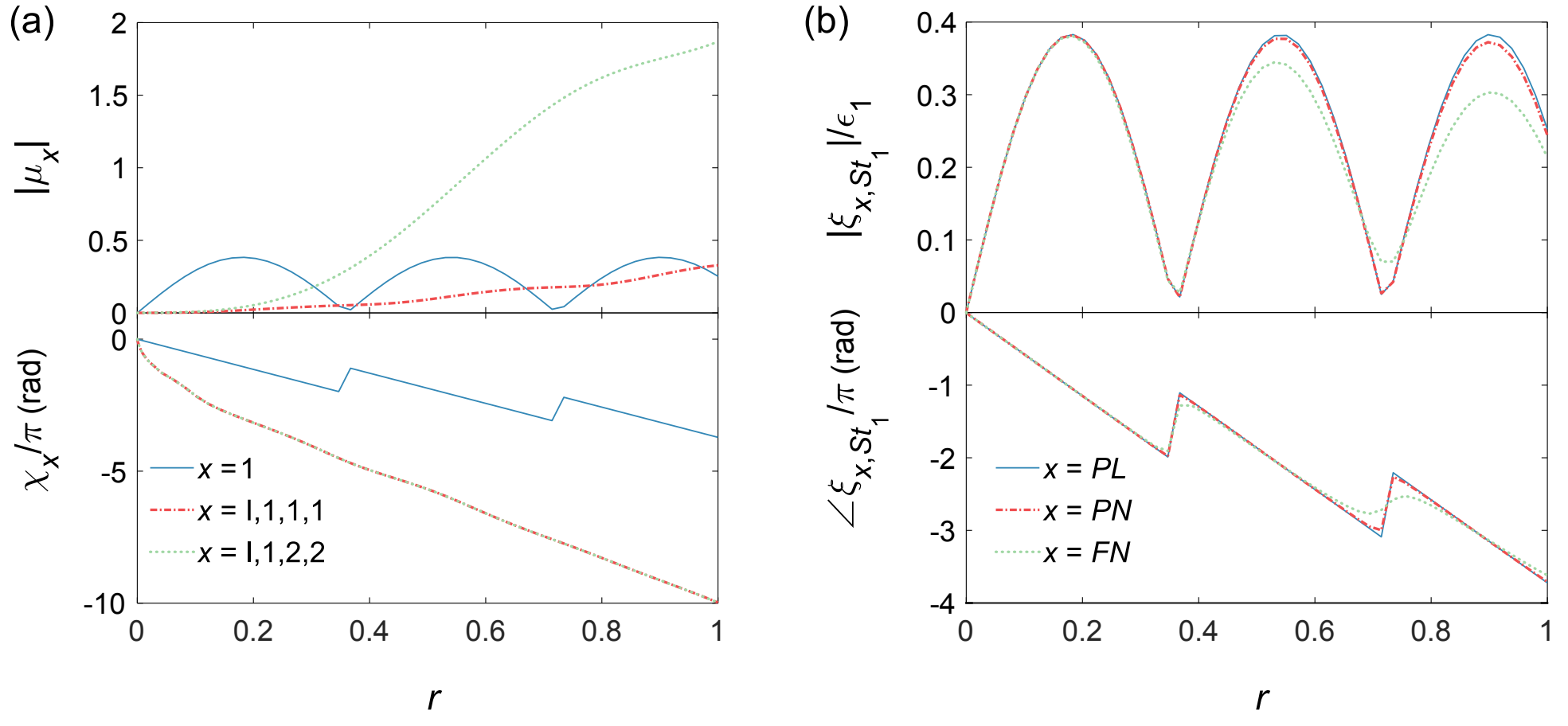

Figure 6. Flame-front response at $St_1$ for the V-flame. (a) Magnitude and phase of the linear, self-nonlinear and mutual-nonlinear contributions in equation (2.23). (b) Magnitude and phase of the resulting combined response at $St_1$. $St_1 = 8$, $St_2 = 5$, $\epsilon_1 = \epsilon_2 = 0.2$, $\cot\alpha = 3$, and $\Omega_b = 0$.

The underlying mechanisms of this phenomenon are analysed below. Figure 6(a) shows the magnitude and phase of each term in equation (2.23). At certain excitation frequencies, the magnitude of the self-nonlinear solution is smaller than that of the mutual-nonlinear solution (this relation depends on the modulation frequency). Their phases are identical ($\chi_{I,1,1,1} = \chi_{I,1,2,2}$), which are insensitive to the modulation frequency, since they both originate from the third-order partial differential equations (PDEs). More interestingly, $\xi_1$, $\xi_{I,1,1,1}$ and $\xi_{I,1,2,2}$ are not affected by the initial phase difference $\delta$ between the two excitations; thus, $\xi_{St_1}$ is also insensitive to $\delta$ (see figure 5(b)). The terms $\xi_1$ and $\xi_{1,1,1}$ can be readily interpreted, as they are influenced solely by the excitation at $St_1$. The latter induces $\xi_{I,1,1,1}$, which is naturally insensitive to $\delta$. $\xi_{1,2,2}$, dominated by excitations at both $St_1$ and $St_2$, produces $\xi_{I,1,2,2}$, which is likewise unaffected by $\delta$. This characteristic has been confirmed by a related experimental study (Zheng *et al.* 2022), supporting the validity of the present modelling results. Furthermore, these results can also be interpreted from another perspective: when $St_1$ and $St_2$ are incommensurate, the initial phase difference disappears after a few cycles and has no effect on the fundamental frequency response. Owing to the above characteristics, the following results are based on $\delta = 0$. Figure 6(b) shows that due to the introduction of nonlinear solutions (both the self-nonlinear and mutual-nonlinear terms, where $\xi_{PL,St_1} = \epsilon_1\xi_1$, $\xi_{PN,St_1} = \epsilon_1\xi_1 + {\epsilon_1}^3\xi_{I,1,1,1}$ and $\xi_{FN,St_1} = \xi_{St_1}$), the magnitude and phase of the flame spatial response at $St_1$ are corrected based on the linear result. Among these, as the downstream distance increases, the amplitude of flame wrinkles decreases, and the phase-step phenomenon observed in the linear results is mitigated.

To compare directly the nonlinear effects of excitations at $St_1$ and $St_2$ on the spatial response of the flame at $St_1$ (which demonstrates the ability to correct the linear results), the relative value $H_{St_1}(r)$ of the mutual-nonlinear and self-nonlinear amplitudes is directly defined as $\mu_{I,1,2,2}(r) - \mu_{I,1,1,1}(r)$, since their corresponding phases are always identical. As shown in figure 7, it is evident that the amplitude of the mutual-nonlinear term is larger than that of the self-nonlinear term, particularly near the flame tip, except within a narrow range of low forcing frequencies. As $St_2$ increases, $\mu_{I,1,2,2}(r)$ exhibits a steady peak, increasing monotonically before decreasing. This peak represents the strongest effect of perturbation at frequency $St_2$ in suppressing the amplitude of flame wrinkles at $St_1$. It is noteworthy that, regardless of $St_1$, the steady peak of the mutual-nonlinear solution always exists and corresponds to the case of $St_2 \approx 3.5$, which is determined by the relation between $St$ and $K$. The nonlinear solution usually varies monotonically with the modulation frequency or $u_c$ ($u_c \in [\bar{u}, c]$, where $c$ is the speed of sound) (Jiang *et al.* 2022). The effect of flame kinematic restoration makes smaller length-scale wrinkles (associated with smaller $u_c$ and larger $St$) more prone to destruction, thereby enhancing the flame kinematic response associated with dissipation. However, $St$ and $u_c$ are negatively correlated (see figure 3), resulting in a stable peak of the nonlinear response.

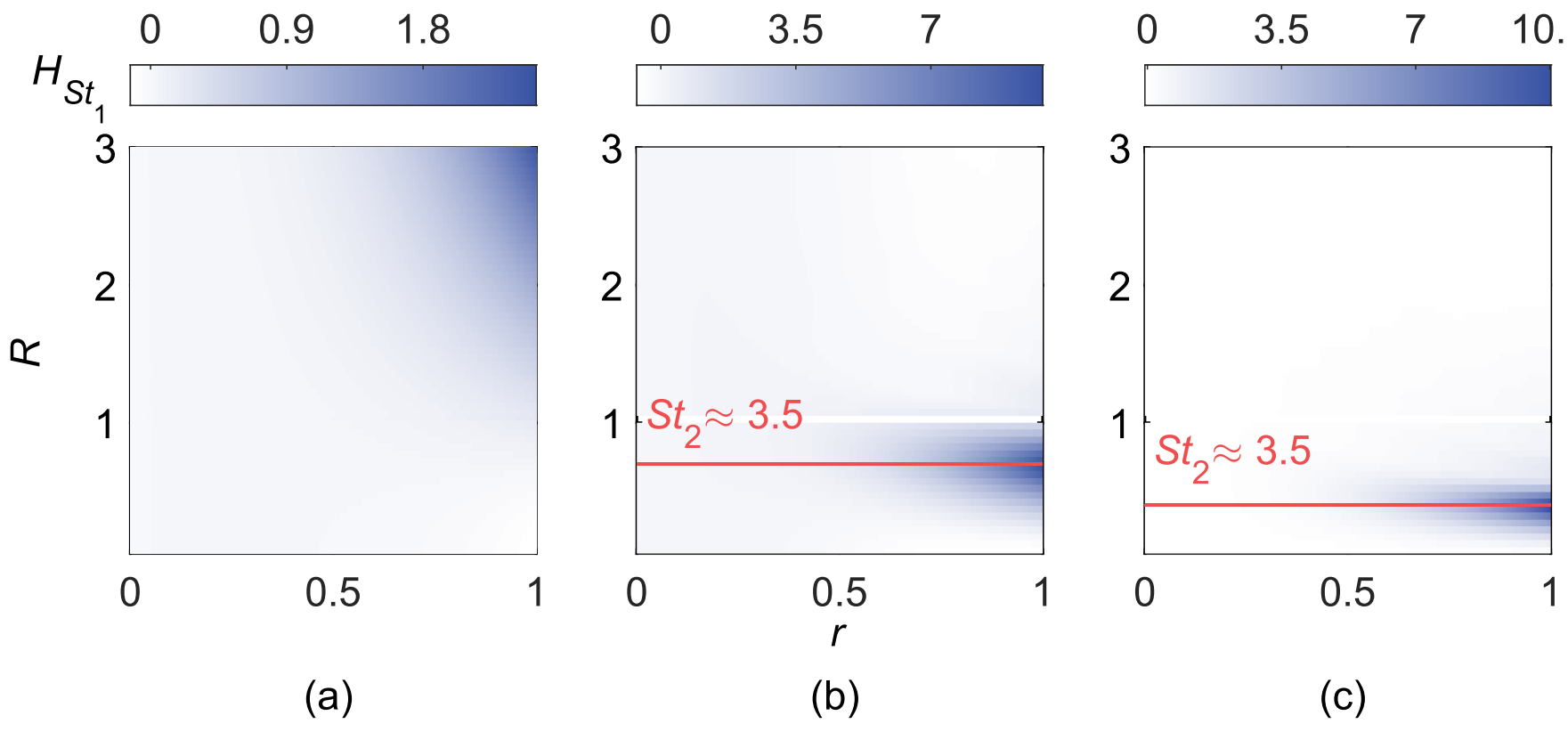

Figure 7. Relative difference between the mutual-nonlinear and self-nonlinear amplitudes for the V-flame. (a) $St_1 = 1$; (b) $St_1 = 5$; (c) $St_1 = 9$. Here, $H_{St_1}(r) = \mu_{I,1,2,2}(r) - \mu_{I,1,1,1}(r)$. $R = St_2/St_1$, $\cot\alpha = 3$, and $\Omega_b = 0$.

## 4. Global nonlinear flame response

The HRR of a premixed flame can be determined by quantifying the perturbation area obtained through flame front-tracking (Schuller *et al.* 2020). However, the presence of excitations alters the time-averaged ensemble kinematics of the flame (Acharya & Lieuwen 2021), which in turn affects the effective displacement speed and HRR of the flame. This mechanism can be further elucidated through asymptotic analysis. Specifically, the terms $\xi_{1,1}$ and $\xi_{2,2}$ in equation (2.19) (see equations (B 14) and (B 15) for their explicit expressions) demonstrate that acoustic forcing generates a zero-frequency component, which modifies the time-averaged flame shape relative to the steady unforced state. The corresponding zero-frequency response will be examined in detail later.

This phenomenon, not previously documented in the literature, suggests that excitation can impact the global response of the flame through an alternative pathway. In particular, the flame displacement speed $S_{\mathrm{eff,disp}}$ employed in HRR calculations differs from that of a steady flame.

Building on the above analysis and equation (2.4), the flame displacement speed accounting for the time-averaged flame dynamics is given by

$$S_{\mathrm{eff,disp}}(r) = \left\langle -\frac{\partial\xi/\partial t - u}{\left[1 + (\partial\xi/\partial r)^2\right]^{1/2}} \right\rangle \tag{4.1}$$

where "⟨⟩" denotes the time-averaged properties. A similar method has been employed in previous studies, e.g., Palies *et al.* (2011). The quantity $\xi$ can be obtained either through the numerical approach or by asymptotic analysis, as given in equation (2.19). Notably, to ensure the accuracy of the results, the numerical approach is generally preferred.

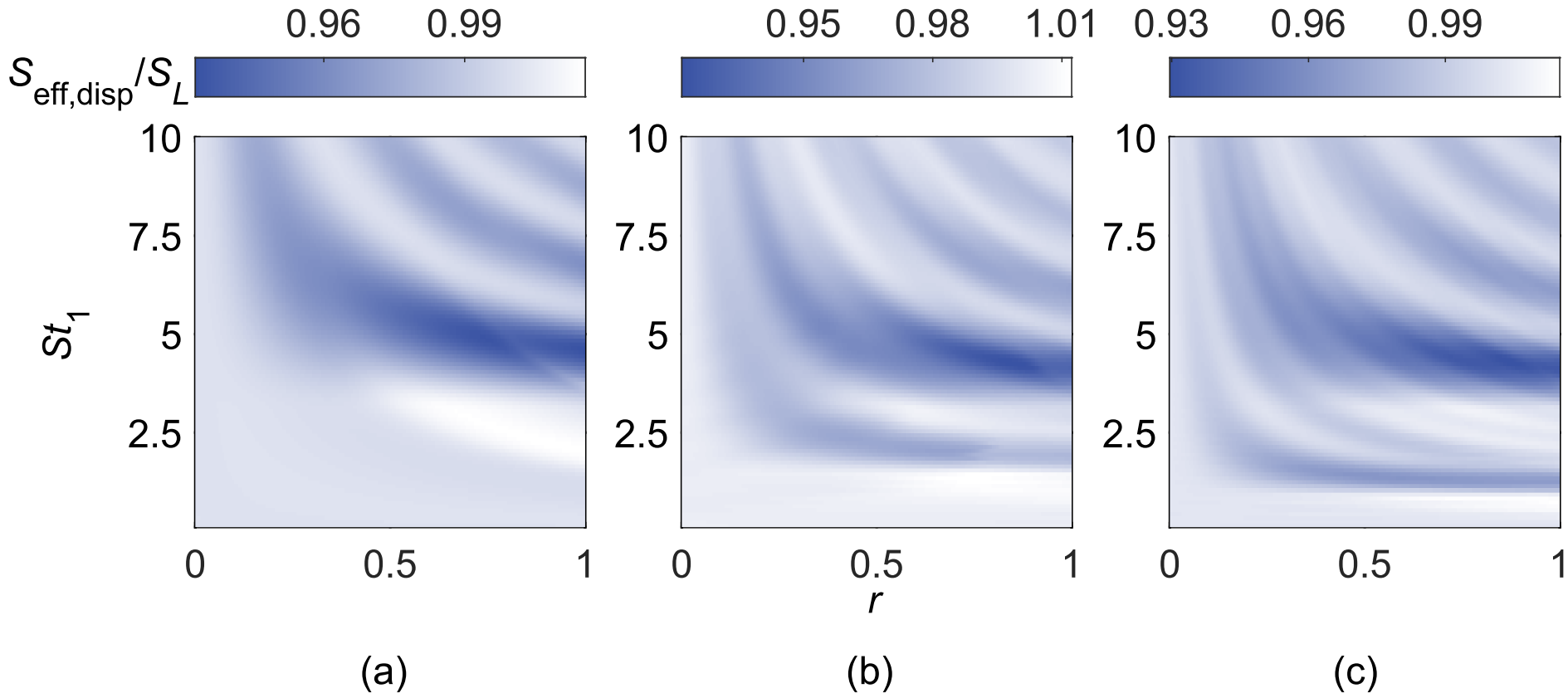

Figure 8. Effective flame displacement speed $S_{\text{eff,disp}}(r)$ for the V-flame. (a) $R = 0$ (single-frequency forcing); (b) $R = 2.2$; (c) $R = 3.6$. Here, $R = St_2/St_1$, $\epsilon_1 = \epsilon_2 = 0.2$, $\cot\alpha = 3$, and $\Omega_b = 0$.

This preference arises because only $\xi_{St_1}$ obtained from asymptotic analysis has been validated against numerical solutions, whereas the general form of $\xi$ has not. Importantly, equation (4.1) is not used to determine the flame position itself, but rather to evaluate the effective flame displacement speed entering the HRR calculation.

Figure 8 shows the variation of $S_{\text{eff,disp}}/S_L$ with spatial position. Due to the disturbances upstream of the flame, the effective displacement speed varies with spatial position and differs from the laminar flame speed $S_L$. In general, $S_{\text{eff,disp}}$ remains consistently lower than $S_L$, indicating a slight suppression of the time-averaged flame front kinematics induced by the excitation. As the flame moves away from the anchor point, the nonlinear kinematics are further amplified, and consequently, the characteristics of the flame speed response to perturbations become increasingly significant. These features are highly sensitive to the harmonic frequency, which governs the nonlinear characteristics of the flame. Furthermore, the introduction of an additional frequency excitation renders the spatial variation of the flame displacement speed more evident.

A correction method for the instantaneous HRR calculation, which accounts for the change in effective flame displacement speed, is as follows:

$$\dot{Q}(t) = \int_{\theta=0}^{2\pi} \int_{r=\Omega_b}^{\Omega} \rho h_R S_{\text{eff,disp}}(r) \left[1 + \left(\frac{\partial \xi(r,t)}{\partial r}\right)^2\right]^{1/2} r\mathrm{d}r\mathrm{d}\theta, \tag{4.2}$$

where, the term $\left[1 + (\partial\xi/\partial r)^2\right]^{1/2}$ is mathematically identical to the left-hand side (LHS) of equation (2.21), but it carries a different physical meaning. It represents the different instantaneous slopes of the flame front at different spatial locations, referred to as the local flame front gradient.

The steady HRR can be determined as follows:

$$\overline{\dot{Q}} = \int_{\theta=0}^{2\pi} \int_{r=\Omega_b}^{\Omega} \rho h_R S_L \left[1 + \left(\frac{\partial \xi_0}{\partial r}\right)^2\right]^{1/2} r\mathrm{d}r\mathrm{d}\theta \tag{4.3}$$

The normalised instantaneous HRR fluctuation is given by

$$q'(t) = \left[\dot{Q}(t) - \overline{\dot{Q}}\right] / \overline{\dot{Q}} \tag{4.4}$$

In this work, perfectly premixed flames are considered, such that perturbations in $\rho$ and $h_R$ can be neglected. Consequently, $\rho$ and $h_R$ vanish in equation (4.4).

The results of asymptotic analysis at $St_1$ are sufficiently accurate to allow the global response at $St_1$ to be quantified by substituting the expression for $\xi_{St_1}$ (equation (2.23)) into equation (4.2), as previously mentioned. However, in certain special cases, the flame response at $|mSt_2 - nSt_1|$ (when $m \neq 0$ and $n \neq 1$) affects that at $St_1$. For instance, when $St_2 = 2St_1$ or $St_2 = St_1$, the response at $|St_2 - St_1|$ or $St_2$ contributes to that at $St_1$. Such inter-frequency coupling is not represented in the single-frequency asymptotic formulation given in equation (2.23). Therefore, the global response at $St_1$ is obtained from numerical results. However, it should be noted that the majority of the response at $St_1$ can be fully and exactly explained by this asymptotic analysis. To achieve a complete and accurate description of the flame response, linearisation of the flame displacement speed and the local flame front gradient is not considered.

Since the HRR is a nonlinear function of the two excitation frequencies, it contains all possible linear combinations of $St_1$ and $St_2$. Using the double Fourier series expansion, $q'$ can be expressed as

$$q' = \sum_m \sum_n \hat{q}_{mSt_1+nSt_2} \cos\left[(mSt_1 + nSt_2)\,t + \varphi_{mSt_1+nSt_2}\right] \tag{4.5}$$

where $m, n \in \mathbb{Z}$; $\hat{q}$ and $\varphi$ represent the amplitudes and phases, respectively, and are functions of perturbation frequency and amplitude.

Consistent with the main focus of this work, only the numerical results corresponding to $St_1$ are extracted from equation (4.5), and the expression for the FDF of two inputs (FDF-TI) is as follows:

$$F_{St_1} = \frac{\hat{q}_{St_1} e^{\mathrm{i}\varphi_{St_1}}}{\epsilon_1} \tag{4.6}$$

As mentioned earlier, the applied threshold $\varLambda = 0.2$ was derived experimentally for a conical flame, so it is necessary to use the experimental FDF results for the V-flame to further validate the accuracy of the flame response to excitation predicted by the current model. To enable comparison with the results of Durox *et al.* (2009), the current dual-input model is first temporarily downgraded to the single-frequency input case, i.e., the excitation corresponding to frequency $St_2$ vanishes, leaving only the excitation at frequency $St_1$. The classical linear results of Schuller *et al.* (2003) are also included for comparison.

As shown in figure 9, it is clear that the present nonlinear model greatly improves the prediction accuracy, especially in terms of FDF gain, compared with the linear results in both the uniform perturbation ($K = 0$) and convective perturbation ($K = 1$) cases. Nonlinear modelling captures key features of the experimental flame response gain to perturbations that were previously unrealistic in the linear case. This improvement arises from two factors: the inclusion of nonlinear effects associated with flame kinematics (namely nonlinear flame restoration, local flame front gradient and effective flame displacement), and the incorporation of the upstream flow profile (i.e., the dependence of $K$ on $St$). To further clarify the respective contributions of these two mechanisms, a sensitivity analysis of $\varLambda$ in the flame response has been carried out.

The results clarify the distinct roles of nonlinear flame kinematics and the $St$-$K$ relation. At low modulation frequencies, $K$ remains equal to unity regardless of $\varLambda$ (see figure 3), indicating that the improved agreement in this regime is primarily due to nonlinear flame kinematics rather than the specific choice of $\varLambda$. As the modulation frequency increases, different values of $\varLambda$ lead to different critical frequencies at which $K$ departs from unity, and the model predictions become increasingly sensitive to $\varLambda$. Notably, the value $\varLambda = 0.2$, calibrated from conical flame data, yields comparatively better agreement with V-flame measurements at higher modulation frequencies than other tested values. While this does not constitute a direct validation, it suggests that $\Lambda = 0.2$ provides a physically reasonable and robust parameter choice within the present modelling framework, and further provides additional confidence in the predictions of the present nonlinear model for further analysis of the two-frequency perturbed input case.

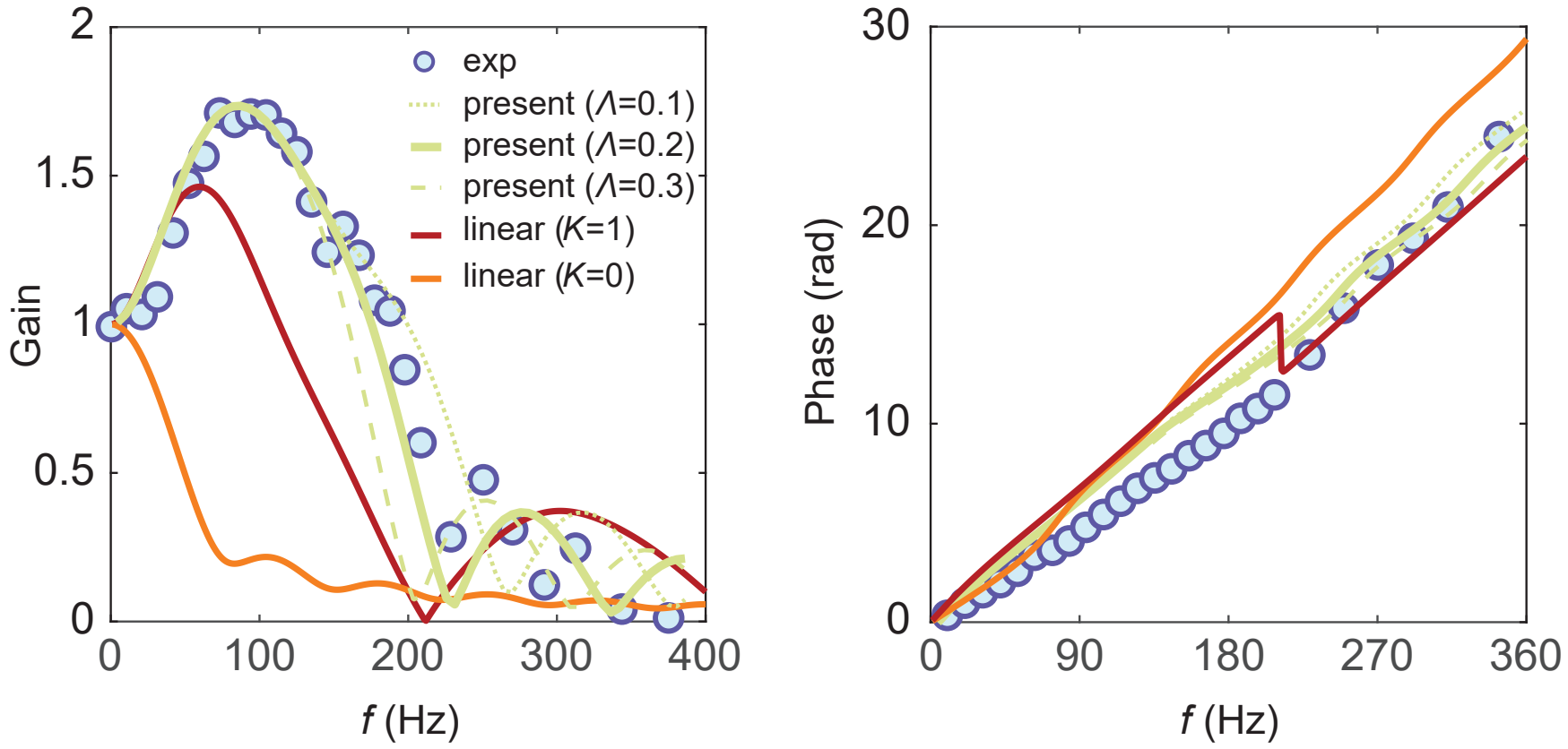

Figure 9. Comparison between linear predictions (FTF), nonlinear predictions (present FDF), and experimental results for the V-flame. The linear results are from Schuller *et al.* (2003), where $K = 1$ corresponds to convective forcing and $K = 0$ to spatially uniform forcing. The nonlinear results are obtained from the present model (equation (4.6)), which incorporates the dependence of $K$ on $St$. Experimental data are from Durox *et al.* (2009). The stabilised flame height is 19.88 mm, the outer burner radius is 17.11 mm, the inner burner radius is 3 mm, the mean bulk velocity is 2.12 m $s^{-1}$, and the normalised perturbation amplitude is $\epsilon = 0.2$.

It is acknowledged that discrepancies between the model predictions and the experimental FDF gain become more pronounced as the forcing frequency increases. Similar non-systematic discrepancies at high forcing frequencies have been reported even in fully nonlinear simulations. For instance, Schuller *et al.* (2003) observed deviations between experiments and nonlinear $G$-equation simulations for conical flames, despite the absence of a $u_c(St)$ formulation. This suggests that such discrepancies are not uniquely attributable to a specific modelling assumption. Instead, experimental studies indicate that increasing forcing frequency promotes the development of small-scale flame wrinkling (Durox *et al.* 2009; Birbaud *et al.* 2006). These fine-scale structures introduce additional, frequency-dependent modifications to the flame surface and thus to the global HRR. In the present model, the HRR is derived from the evolution of the global flame surface under harmonic forcing, which captures the large-scale dynamics but does not explicitly resolve these small-scale wrinkling effects. Therefore, at higher $St$, the observed discrepancies are more likely associated with the increasing importance of the unresolved fine-scale flame dynamics, rather than a single missing modelling ingredient.

Although the overall agreement is satisfactory, a phase discrepancy remains in figure 9(b), particularly in the moderate-frequency range. The observed phase discrepancy primarily originates from a mismatch in the convective time scale used in the model. In the present framework, the phase of the flame response is largely governed by the convective delay between the velocity perturbation and the flame response, which scales with the ratio of a characteristic length to the mean upstream velocity $\bar{u}$. As a result, an underestimation of $\bar{u}$ directly leads to an overprediction of the phase lag. In the experiments of Durox *et al.* (2009), the bulk velocity at the nozzle outlet, $\bar{u}_0 = 2.12$ m $s^{-1}$, was reported and used in the present model as a proxy for the upstream mean velocity. However, PIV measurements reported by Birbaud *et al.* (2006) indicate that the local velocity in the unburned region upstream of the flame is generally higher than the bulk outlet value. Therefore, using $\bar{u}_0$ as the convective velocity leads to an overestimation of the convective time and consequently to a phase shift. To assess the sensitivity of the phase prediction to this parameter, a consistency check was performed by varying the upstream mean velocity. The results indicate that a moderately higher convective velocity improves the agreement of the phase response in the moderate-frequency range, which is consistent with experimental observations suggesting that the local upstream velocity ahead of the flame may exceed the bulk outlet velocity reported in Durox *et al.* (2009). Although variations in $\bar{u}$ affect the frequency mapping through the definition of the Strouhal number, the gain response predicted by the present model remains in satisfactory agreement with the experimental measurements. Overall, the model captures the dominant gain and phase trends over the investigated frequency range.

Figure 10 shows the dependence of the FDF-TI of the V-flame on dual-frequency excitation. The results depicted in figure 10(a) demonstrate that the gain of the flame response exhibits typical characteristics of a V-flame, where the FDF-TI gain exceeds unity within the low-excitation-frequency range, indicating that the flame acts as an amplifier of disturbances. When the disturbance at $St_2$ is considered, the features with gains greater than 1 are significantly altered and vary with the frequency $St_2$, particularly for $R \in (1, 2]$, where the single peak splits into multiple peaks. There are even three local maxima when $St_2$ is twice as large as $St_1$. In general, in non-experimental studies, the V-flame has a single peak with a value above 1 (Schuller *et al.* 2020), with the corresponding frequency considered the “resonant frequency” of the flame (Kedia *et al.* 2011). The flame gain exceeding unity plays a key role not only in the potential for intrinsic thermoacoustic (ITA) instability (Xu *et al.* 2021; Silva 2023; Emmert *et al.* 2015), but also in analysing the mechanisms of thermoacoustic instability formation.

Figure 10(b) further illustrates the role of the perturbation at $St_2$ in the FDF-TI by introducing the parameter $T_{\mathrm{V},St_1}$. The value of $T_{\mathrm{V},St_1}$ is given by $\left|F^{PN}_{\mathrm{V},St_1}\right| - \left|F^{FN}_{\mathrm{V},St_1}\right|$, where the superscript “$FN$” represents the full nonlinear solution, including all nonlinearities caused by the two input excitations (the cases of $R \neq 0$). The superscript “$PN$” represents the partial nonlinear solution, in which all nonlinearities induced by perturbation at $St_2$ vanish (the case where a dual-frequency input flame degenerates into a single-frequency perturbed-input flame, i.e., $R = 0$). A positive value of $T_{\mathrm{V},St_1}$ indicates that the perturbation at $St_2$ weakens the magnitude of the flame response. Two boundary lines define the region within which the difference value $T_{\mathrm{V},St_1}$ is significant, whereas outside this region it is negligible, except for certain special cases. The left boundary is the frequency threshold for the linear flame response, and the right boundary is the frequency threshold for saturation of the nonlinear flame response.

Low-order asymptotic analysis shows that the $\xi_{\mathrm{I},1,2,2}$ term in equation (2.23) primarily accounts for the properties of the FDF-TI gain. When the modulation frequency is sufficiently small ($St < 0.9$), the effect of $\xi_{\mathrm{I},1,2,2}$ is suppressed, and the gain is insensitive to the perturbation at $St_2$. With increasing $St_2$, a distinctive nonlinear response feature emerges, whereby $\xi_{\mathrm{I},1,2,2}$ significantly modifies the gain characteristics of the FDF-TI. Similar to the discussion in figure 7, the negative correlation between $St$ and $K$ dominates the value of $\xi_{\mathrm{I},1,2,2}$, giving rise to a critical frequency ($St_2 = 4.3$) for the flame response gain. Beyond this frequency, the influence of the perturbation at $St_2$ on the gain diminishes rapidly, and the flame enters the saturation region of the response.

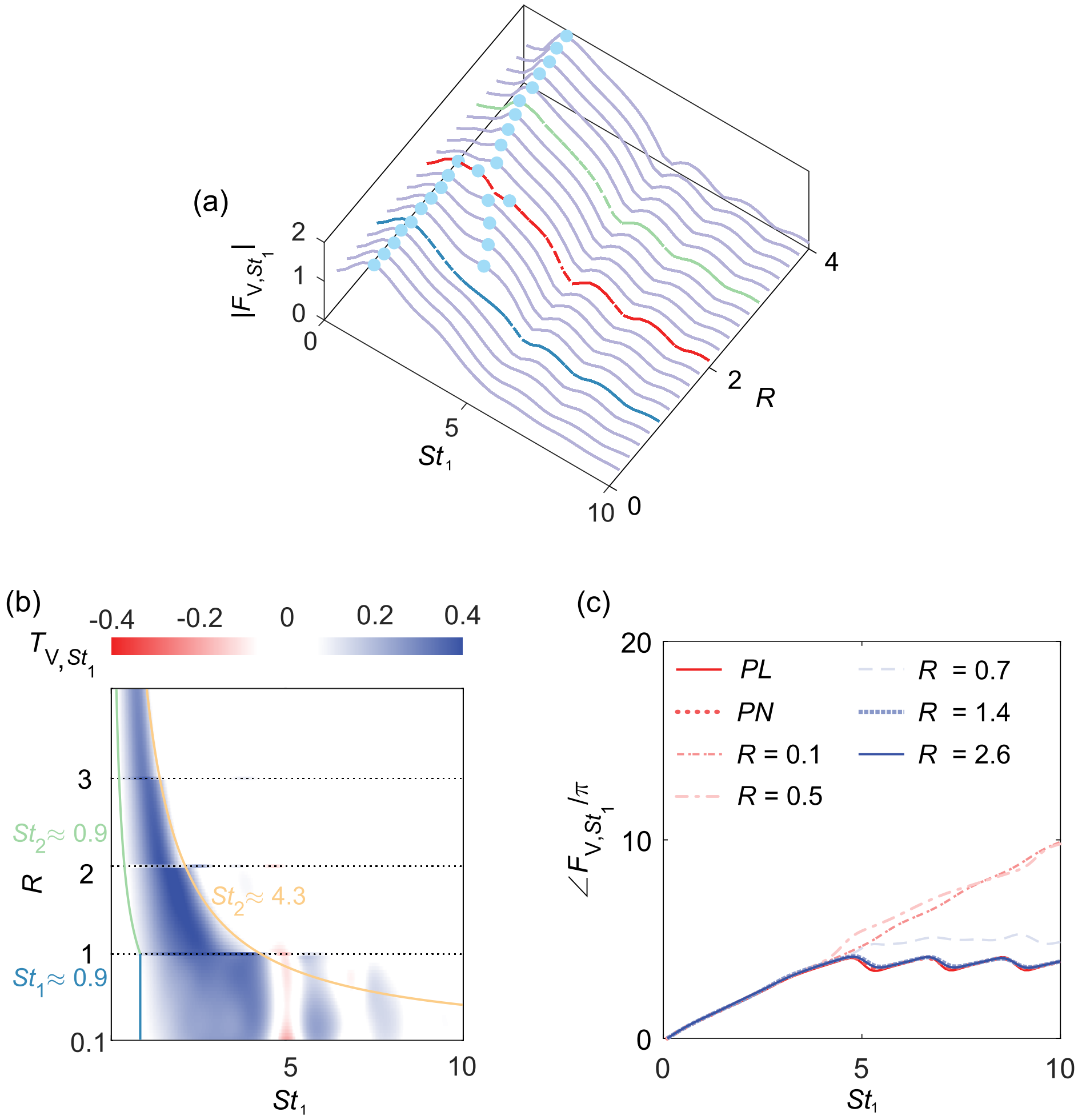

Figure 10. The FDF-TI response of the V-flame under dual-frequency forcing. (a) Dependence of the FDF-TI gain at $St_1$ on the two forcing frequencies. Symbols "∘" denote local maxima with gain values greater than unity (referred to as "peaks"). (b) Influence of the two forcing frequencies on the relative value $T_{V,St_1}$. (c) Dependence of the FDF-TI phase on the forcing frequencies in selected cases. Here, "*PL*" denotes the purely linear solution, where nonlinearities associated with flame kinematic restoration, flame displacement speed, and local flame-front gradient are neglected, and "*PN*" denotes the partially nonlinear solution, where nonlinearities induced by perturbation at $St_2$ are neglected. Here, $\epsilon_1 = 0.2$, $\epsilon_2 = 0.1$, $R = St_2/St_1$, $\cot\alpha = 3$, and $\Omega_b = 0$.

In some special cases, the effect of the perturbation at $St_2$ on the FDF-TI outside the critical-frequency boundary is evident, especially for small $R$. In fact, whenever $R$ transitions to a natural number (e.g., 1, 2, 3), there is a noticeable jump in the value of $T_V$, as some additional gain appears in these special cases. These so-called "errors" occur because the flame response at $|mSt_2 - nSt_1|$ (when $m \neq 0$ and $n \neq 1$) affects that at $St_1$. For instance, when $St_2 = 2/3St_1$ or $St_2 = 2St_1$, the response at $|3St_2 - St_1|$ or $|St_2 - St_1|$ contributes to that at $St_1$. These new contributions are not included within the scope of the original flame response at $St_1$, and therefore cannot be described by the corresponding asymptotic analysis.

In figure 10(c), it can be observed that the phase changes exhibit more regularity than the complex behaviours of the FDF-TI gain. The influence of the perturbation at $St_2$ is confined to the high-frequency range of $St_1$ and is only noticeable when $St_2 < St_1$. This is attributed to the inverse relation between $St$ and $K$, resulting in $K_2 > K_1$ when $St_2 < St_1$. As $K$ increases, $u_c$ decreases steadily, which restricts the propagation of acoustic disturbance and amplifies the response of the FDF-TI phase to excitations.

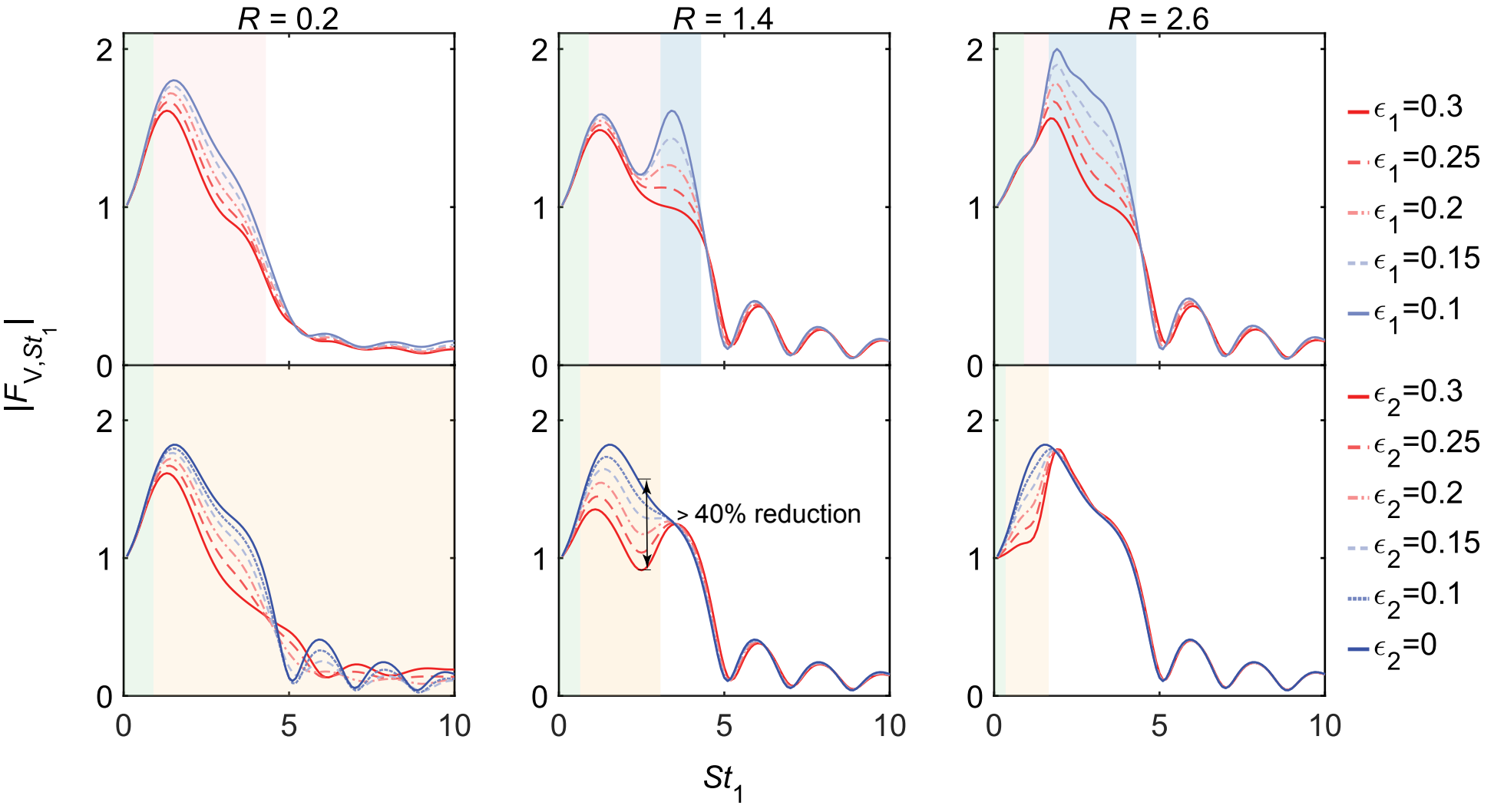

Figure 11. Dependence of the FDF-TI gain on the two forcing amplitudes. The top panels show cases with $\epsilon_2 = 0.2$, whereas the bottom panels show cases with $\epsilon_1 = 0.2$. The right boundary of the green region corresponds to cases where either $St_1$ or $St_2$ is approximately 0.9. In contrast, the right boundary of the red region, or that of the blue region when present, corresponds to $St_1 \approx 4.3$, whereas the right boundary of the orange region, when identifiable, corresponds to $St_2 \approx 4.3$.

By analysing the amplitude variation of the two excitations, as shown in figure 11, the nonlinear gain characteristics can be further elucidated. The destructive features become more pronounced with higher perturbation amplitude. Particularly, when subjected to a perturbation at $St_1$ with $\epsilon_1 = 0.2$, the HRR gain at $St_1$ decreases by more than 40% compared with the single-frequency input after introducing a perturbation at $St_2$ with $\epsilon_2 = 0.3$ (see the case of $R = 1.4$). This occurs because large-amplitude perturbations tend to generate small-scale wrinkles, which are readily smoothed out by flame motion, thereby reducing the gain.

However, it should be emphasised that a reduction of the HRR gain at $St_1$ alone does not provide a sufficiently comprehensive assessment of the control effect. Under dual-frequency excitation, the HRR response is inherently nonlinear, and its spectral content extends over a range of combination frequencies $|mSt_1 + nSt_2|$. Consequently, a reduction at a single fundamental component does not necessarily imply a reduction in the overall thermoacoustic driving, as energy may be redistributed across the spectrum. A detailed analysis of all spectral components of the fluctuating HRR for this representative case is therefore presented later.

The above behaviour is confined to the effective frequency range bounded by the two previously identified limits (0.9 and 4.3). In addition, the green-shaded regions correspond to the linear case, i.e., $St < 0.9$. The effective region can be used to explain the mechanism of multi-peak formation of the FDF-TI gain.

For the single-peak cases ($R < 1$ and $R > 2$), the behaviours of the FDF-TI gain are similar, but the underlying mechanism is different. When $R < 1$ (here $R = 0.2$), the peak lies within the effective frequency range of both $St_1$ and $St_2$. Consequently, its behaviour is sensitive to the amplitudes of both excitations (with the $St_1$ case shown in red and the $St_2$ case shown in yellow). Conversely, when $R > 2$ (here $R = 2.6$), the peak falls within the effective frequency range of $St_1$ but lies outside the primary effective range of $St_2$. As a result, it is predominantly controlled by the $St_1$ disturbance, shown in blue in the figure ($4.3/R < St_1 < 4.3$). However, in the remaining effective region for the $St_1$ case, i.e., $0.9 < St_1 < 4.3/R$ (shown in red), the influence of $St_1$ on the FDF-TI gain becomes relatively weaker. This is because the effective region associated with $St_2$ (shown in yellow) continues to contribute to the reduction of the FDF-TI gain, thereby causing the nonlinear flame response to exhibit saturation with respect to the disturbance amplitude at $St_1$.

For the double-peak scenario ($1 < R < 2$), considering $R = 1.4$ here, the first peak lies within the effective region of $St_2$ and is primarily governed by $St_2$. Although influenced by $St_1$, the flame response to the disturbance amplitude tends towards saturation, similar to the red region observed at $R = 2.6$, thereby weakening the influence of $St_1$. The second peak is predominantly located within the blue region, thus being entirely controlled by the $St_1$ disturbance. Its behavioural characteristics again correspond to the phenomena observed at $R = 2.6$.

In the triple-peak case ($R = 2$, see figure 10(a)), the first two peaks form for similar reasons to the double-peak case, with the third peak clearly associated with the frequency $|St_2 - St_1|$. When $R = 2$, $|St_2 - St_1|$ becomes $St_1$, contributing to the additional gain in flame response at $St_1$.

To provide a more comprehensive assessment, all spectral components of the fluctuating HRR have been examined for the representative case in which the reduction in HRR gain at $St_1$ exceeds 40%. Figure 12 presents a comparison of the normalised HRR spectral amplitudes at different frequency components under single- and dual-frequency excitation. These results are derived from equation (4.5). Spectral amplitudes involving both $St_1$ and $St_2$ are normalised by $\sqrt{\epsilon_1^2 + \epsilon_2^2}$, whereas those associated with a single excitation frequency are normalised using the corresponding perturbation amplitude. As indicated by the asymptotic analysis, the zero-frequency component arises independently and must therefore be treated separately. Under dual-frequency excitation, it originates from both the terms at $St_1 - St_1$ and $St_2 - St_2$, whereas under single-frequency excitation it arises solely from the term at $St_1 - St_1$. Accordingly, the zero-frequency response is first decomposed according to its origin and subsequently normalised by the corresponding excitation amplitudes.

Even under single-frequency excitation, the nonlinear flame dynamics generate spectral components other than $St_1$, as reported in the experiments of Albayrak *et al.* (2018), although their amplitudes remain small compared with that of the fundamental component. After introducing excitation at frequency $St_2$, the previously observed spectral components associated with $St_1$ are suppressed, and in several cases the reduction at these components is more pronounced than that observed at $St_1$ itself. The only exception is the zero-frequency component, for which the contributions from $St_1$ and $St_2$ involve no relative phase difference and therefore combine through direct linear superposition.

While some spectral components, particularly those associated with $St_2$, are enhanced under dual-frequency excitation, it is essential to evaluate whether this redistribution is accompanied by an overall decrease in the total fluctuating HRR energy. Based on the energy definition $E = \left\langle q'^2(t) \right\rangle = A_0^2 + \sum_{St_i \neq 0} \frac{A_{St_i}^2}{2}$, where the fluctuating HRR is defined relative to the steady-state flame (see equation (4.4)), and the zero-frequency term represents an offset from the steady-state HRR rather than an oscillatory component and is therefore treated separately in the energy evaluation, the ratio of the total energy under dual-frequency excitation to that under single-frequency excitation is approximately 0.75. This indicates that the total fluctuating HRR energy is reduced by approximately 25%. The observed suppression therefore represents not only a redistribution of spectral energy but also a genuine net decrease in the overall fluctuation level.

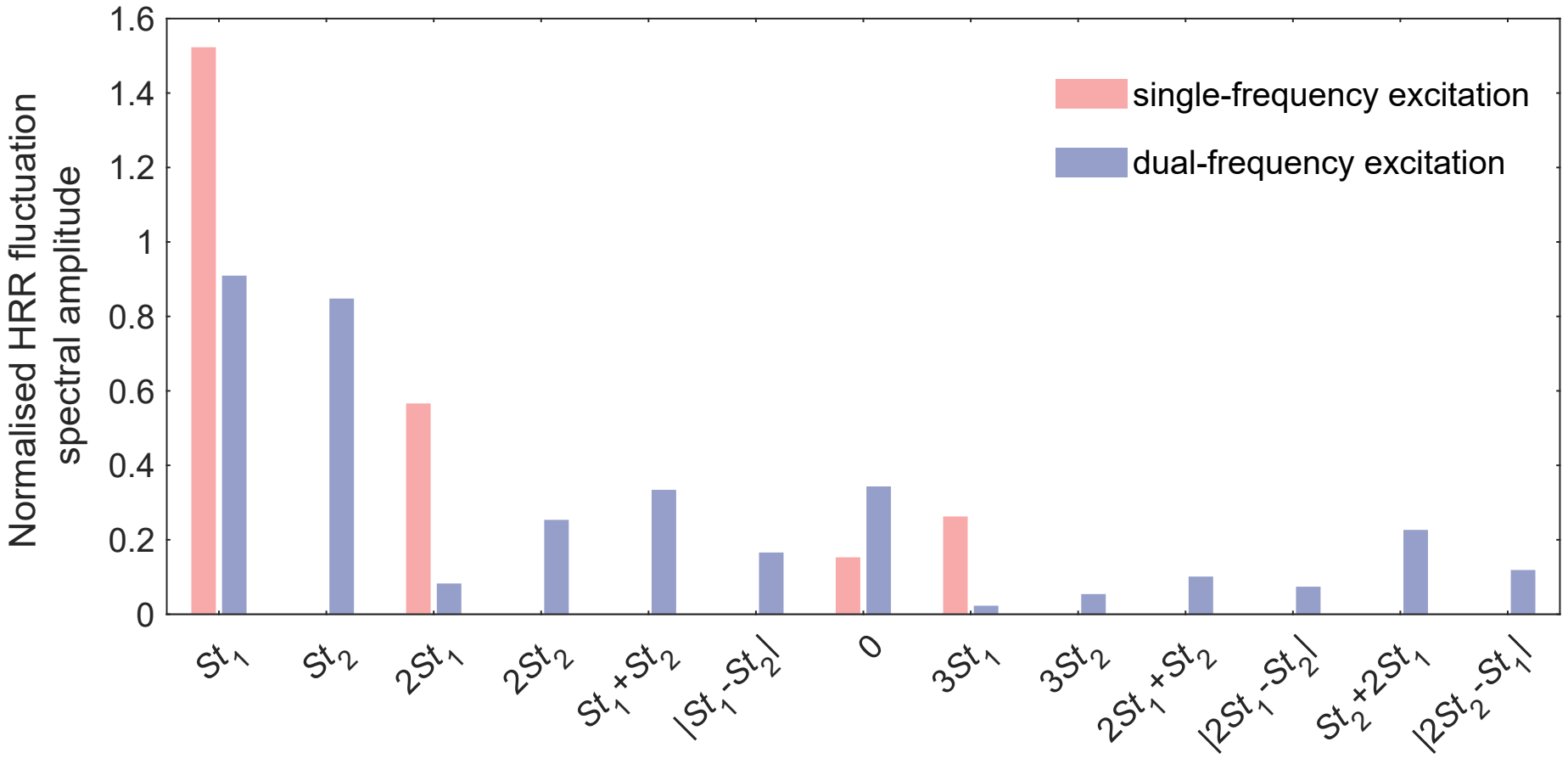

Figure 12. Comparison of the normalised HRR spectral amplitudes at different frequency components under single- and dual-frequency excitation. For the single-frequency case, forcing at $St_2$ is absent. For the dual-frequency case, $St_2 = 3.5$ and $\epsilon_2 = 0.3$. Here, $\cot\alpha = 3$, $St_1 = 2.5$, $\epsilon_1 = 0.2$, and $\Omega_b = 0$.

However, practical considerations must also be taken into account. Although the total fluctuating HRR energy is reduced, introducing $St_2$ inevitably generates additional spectral components. While their amplitudes are relatively small, their presence cannot be entirely disregarded. In practical combustors characterised by broadband disturbances and multiple unstable modes, newly introduced frequency components that coincide with sensitive system frequencies may trigger or amplify other instabilities. Moreover, the effectiveness of the proposed strategy is confined to a specific frequency range. Minor detuning may weaken the intended suppression effect or even destabilise other modes. Consequently, the selection of $St_2$ must be made with careful consideration of the relevant effective frequency range and the characteristics of the specific combustion system.

## 5. Conclusion

This paper examines the nonlinear response characteristics of laminar V-flames under simultaneous excitation at two distinct frequencies, denoted $St_1$ and $St_2$. The nonlinear features of the spatial and global flame responses were obtained through a low-order asymptotic analysis based on the $G$-equation model, which retained terms up to the third order in the normalised excitation amplitude. Within the modelling framework, the velocity potential function explicitly includes the frequency dependence of the disturbance propagation speed. To ensure the accuracy of the analytical solution and complement it, numerical flame-front-tracking methods based on the $G$-equation model were used. The flame response results are consistent with existing experimental data, validating the effectiveness of the proposed model.

An explicit mechanism by which the perturbation at $St_2$ suppresses the flame kinematics at $St_1$ is shown. From an asymptotic analytical point of view, the third-order mutual nonlinear term induced by the perturbations at $St_1$ and $St_2$ attenuates the folds in the flame-front tracking compared with the single-frequency response. In particular, as $St_2$ increases, there is a steady peak in the mutual nonlinear effect on the spatial response, first increasing monotonically and then decreasing. The reason is that the nonlinear results typically vary consistently with the modulation frequency $St$ or the propagation speed of the flow disturbance $u_c$. The presence of flame kinematic restoration results in a higher likelihood of wrinkles on smaller length scales (associated with smaller $u_c$ and larger $St$) being eliminated, resulting in an increase in the dissipative flame kinematic response. Thus, the positive correlation between $u_c$ and $St$ counteracts the enhanced dissipation associated with increasing $St$, giving rise to the non-monotonic behaviour described above.

Based on the results of the flame spatial wrinkling, the heat release rate (HRR) response is calculated and the two boundaries are identified, thus extracting the effective region where the perturbation at $St_2$ plays an important role in the HRR at $St_1$. The left boundary line represents the frequency threshold in the linear response, while the right boundary line, dominated by the correlation between $u_c$ and $St$, indicates the frequency threshold at which the nonlinear response of the flame saturates. When $St_2$ is located within this effective region, it significantly alters the character of the HRR gain, including features exceeding unity, which is critical not only for potential intrinsic thermoacoustic (ITA) instabilities, but also for analysing the formation of thermoacoustic instabilities.

In addition, increasing the amplitude of the perturbation at $St_2$ greatly disrupts the gain of the nonlinear response. This is because the enhancement of large-amplitude perturbations can easily produce small-scale wrinkles that are readily smoothed out by the flame kinematics, thus reducing the gain. Particularly, when subjected to a perturbation at $St_1$ with a normalised amplitude of 0.2, the HRR gain at $St_1$ decreases by more than 40% compared with a single-frequency input after introducing a perturbation at $St_2$ with a normalised amplitude of 0.3. In terms of the integrated HRR fluctuation energy, dual-frequency excitation results in an approximately 25% reduction compared with the single-frequency case. This suggests that by selectively introducing a perturbation at a non-critical frequency, with careful consideration of the relevant effective frequency range and the characteristics of the specific combustion system, the nonlinearity due to frequency interactions can be harnessed to suppress the flame response at the critical frequency. While the results demonstrate that nonlinear frequency interactions can substantially attenuate the flame response at the primary frequency, such attenuation does not necessarily imply suppression of thermoacoustic instability, as the overall growth or decay of oscillations is governed by the Rayleigh criterion, which depends on both the amplitude and the phase relationship between pressure and HRR fluctuations. A reduction in flame response magnitude alone is therefore insufficient to ensure stability if the relative phase remains within the energy-producing region.

The present framework provides a theoretical foundation for analysing flame dynamics under harmonic forcing; however, its extension to practically relevant turbulent premixed swirling flames, especially in gas-turbine applications, warrants further consideration. The derivation of the present $St$-$u_c$ relation is based on a velocity-potential formulation and therefore assumes an irrotational base flow. Consequently, the model is formally restricted to this framework and cannot be directly applied to rotational flows without substantial reformulation. Extending the present approach to swirling configurations would require a generalised treatment that consistently accounts for mean vorticity and rotational effects, which is beyond the scope of the current study.

In addition to this fundamental limitation, further challenges arise from both swirl and turbulence. For swirling flows, a previous study (Acharya & Lieuwen 2020) has shown that swirl can be incorporated within the $G$-equation framework through azimuthal velocity components and helical modal decomposition. Under simplifying assumptions (e.g., axisymmetric base flames), swirl may not directly alter the steady flame-angle relation, but it significantly modifies disturbance propagation via azimuthal convection and modal coupling. In more general swirling configurations involving recirculation, streamline curvature, and non-uniform velocity distributions, the base-state flame geometry and angle are expected to be altered, requiring reformulation of the underlying geometric relations.

For turbulent flames, additional complexity arises from multi-scale flame wrinkling and turbulence-flame interactions. Within the $G$-equation framework, turbulence effects are typically incorporated either through an effective turbulent flame speed or through fluctuating velocity fields. While these approaches can capture certain aspects of turbulence-flame coupling, they rely on modelling assumptions or perturbation frameworks that are generally limited to weakly or moderately turbulent conditions. Importantly, the characteristic multi-scale wrinkling introduced by turbulence is not explicitly resolved in the present surface-based harmonic formulation, which may limit its applicability at higher turbulence intensities.

Overall, while the present framework provides a useful theoretical basis, its extension to practically relevant turbulent premixed swirling flames requires substantial additional developments to account for rotational effects, modified flame geometry, and multi-scale turbulence-flame interactions.

**Declaration of interests.** The authors report no conflicts of interest.

**Author contributions.** X.J. and X.L. contributed equally to this work. J.L. and X.J. conceived the project. X.J. performed most of the theoretical derivation, simulations, and analyses, to which J.N., J.L., X.L., A.S.M., L.L., T.L., and L.Y. also contributed. X.J., X.L., A.S.M., J.N. and J.L. wrote the manuscript with input from all other authors.

**Acknowledgments.** The authors gratefully acknowledge the anonymous reviewers for their insightful and constructive comments, which have helped improve the clarity and quality of the work. X.J. thanks He Yu, Meizhu Zhou, Haimin Jiang, and Xiaoyun Jiang for their continuous support and encouragement.

**Funding.** This work has been funded by the Chinese National Natural Science Fund for the National Natural Science Foundation of China (Grant No. 52376089) and the European Research Council grant AFIRMATIVE (2018–2023, Grant No. 772080).

## Appendix A. Data processing for local perturbation propagation speed

Birbaud *et al.* (2006) investigated the phase difference $\phi$ between the excitation signal and the axial velocity response of a conical flame subjected to modulated forcing. Figure 13 presents the distribution of $\phi$ along the centreline ($r = 0$) for three different excitation frequencies $f$. Based on this, the local perturbation velocity at $r = 0$ can be expressed as

$$u_c(x) = 2\pi f \frac{dx}{d\phi}. \tag{A 1}$$

It is important to note that the mean convective velocity does not necessarily coincide with the prescribed bulk velocity upstream, as acoustic waves can modify the time-averaged flow velocity (Acharya & Lieuwen 2021). To address this, we adopt a simplified assumption: the perturbation wave propagates at a velocity bounded between two limiting cases, either at the speed of sound (dominated by acoustic effects) or at the mean convective velocity (dominated by hydrodynamic effects). Accordingly, the minimum propagation speed of disturbances is taken to be the mean convective velocity. On this basis, the ratio of local perturbation propagation velocities is defined as

$$\kappa = \frac{\min |u_c(x)|}{|u_c(x)|}, \tag{A 2}$$

and the corresponding results are presented in figure 2.

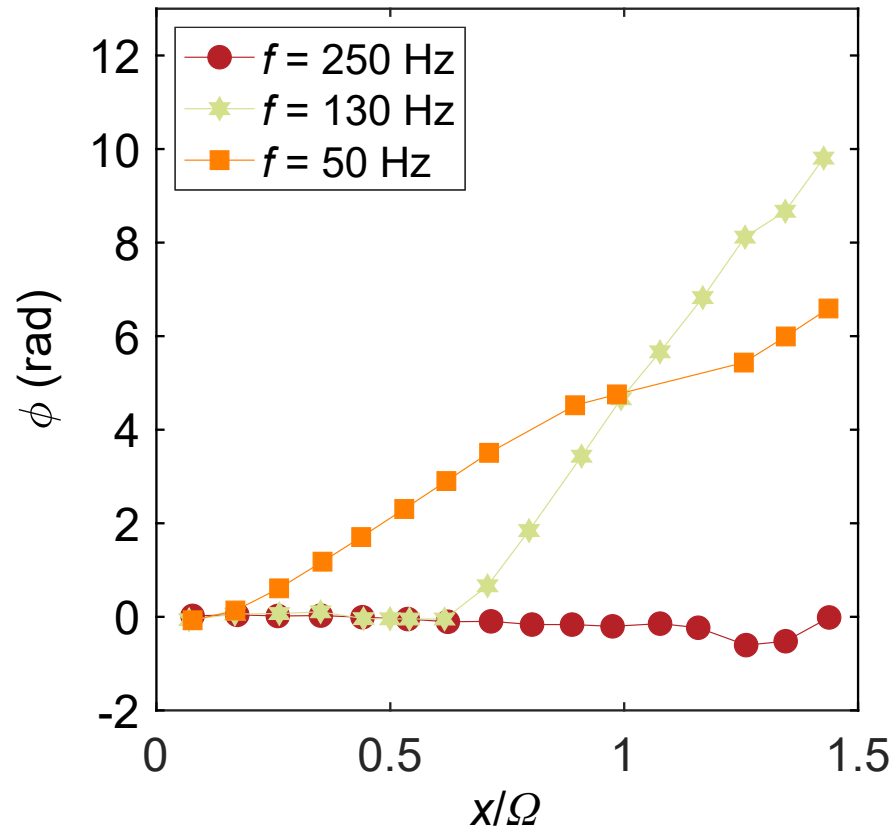

Figure 13. Phase difference between the excitation signal and the axial velocity signal obtained from the experiments of Birbaud *et al.* (2006).

## Appendix B. Derivation of asymptotic solutions

The asymptotic expansion equation (2.19) and the Taylor expansion equation (2.21) can be applied to the dynamic flame-front equation (2.4), and terms of the same order can then be matched. The PDEs of different flames have similar forms. The governing equations at each order for the V-flame front are given below. For the linear terms, they are

$$\frac{\partial \xi_1}{\partial t} - \frac{1}{2}\sin 2\alpha \frac{\partial \xi_1}{\partial r} = C_1(r,t) \tag{B 1}$$

$$\frac{\partial \xi_2}{\partial t} - \frac{1}{2}\sin 2\alpha \frac{\partial \xi_2}{\partial r} = C_2(r,t) \tag{B 2}$$

The self-nonlinear solutions are

$$\frac{\partial \xi_{1,1}}{\partial t} - \frac{1}{2}\sin 2\alpha \frac{\partial \xi_{1,1}}{\partial r} = -\frac{1}{2}\sin^4\alpha \left(\frac{\partial \xi_1}{\partial r}\right)^2 \tag{B 3}$$

$$\frac{\partial \xi_{2,2}}{\partial t} - \frac{1}{2}\sin 2\alpha \frac{\partial \xi_{2,2}}{\partial r} = -\frac{1}{2}\sin^4\alpha \left(\frac{\partial \xi_2}{\partial r}\right)^2 \tag{B 4}$$

$$\frac{\partial \xi_{1,1,1}}{\partial t} - \frac{1}{2}\sin 2\alpha \frac{\partial \xi_{1,1,1}}{\partial r} = -\sin^4\alpha \frac{\partial \xi_1}{\partial r}\frac{\partial \xi_{1,1}}{\partial r} \tag{B 5}$$

$$\frac{\partial \xi_{2,2,2}}{\partial t} - \frac{1}{2}\sin 2\alpha \frac{\partial \xi_{2,2,2}}{\partial r} = -\sin^4\alpha \frac{\partial \xi_2}{\partial r}\frac{\partial \xi_{2,2}}{\partial r} \tag{B 6}$$

and the three mutual-nonlinear solutions are

$$\frac{\partial \xi_{1,2}}{\partial t} - \frac{1}{2}\sin 2\alpha \frac{\partial \xi_{1,2}}{\partial r} = -\sin^4\alpha \frac{\partial \xi_1}{\partial r}\frac{\partial \xi_2}{\partial r} \tag{B 7}$$

$$\frac{\partial \xi_{1,2,2}}{\partial t} - \frac{1}{2}\sin 2\alpha \frac{\partial \xi_{1,2,2}}{\partial r} = -\sin^4\alpha \left(\frac{\partial \xi_1}{\partial r}\frac{\partial \xi_{2,2}}{\partial r} + \frac{\partial \xi_2}{\partial r}\frac{\partial \xi_{1,2}}{\partial r}\right) \tag{B 8}$$

$$\frac{\partial \xi_{1,1,2}}{\partial t} - \frac{1}{2}\sin 2\alpha \frac{\partial \xi_{1,1,2}}{\partial r} = -\sin^4\alpha \left(\frac{\partial \xi_1}{\partial r}\frac{\partial \xi_{1,2}}{\partial r} + \frac{\partial \xi_2}{\partial r}\frac{\partial \xi_{1,1}}{\partial r}\right) \tag{B 9}$$

These equations are solved sequentially based on the boundary condition equation (2.5). For the linear terms, these are

$$\xi_1(r,t) = \frac{\sin\left[St_1\left(-1+r+\Gamma t\right)/\Gamma\right] - \sin\left[St_1\left(\cot\alpha K_1\left(-1+r\right)+t\right)\right]}{St_1\left(-1+\Gamma\cot\alpha K_1\right)} \tag{B 10}$$

where $\mathit{\Gamma}=\cos\alpha\sin\alpha$, and

$$\xi_2(r,t,\delta)=\frac{\sin\left[\delta-St_2\left(\cot\alpha K_2(-1+r)+t\right)\right]-\sin\left[St_2\left(1+\mathit{\Gamma}\delta-r-\mathit{\Gamma}t\right)/\mathit{\Gamma}\right]}{St_2\left(-1+\mathit{\Gamma}\cot\alpha K_2\right)} \tag{B 11}$$

The explicit solutions for the remaining terms are too complex and, to improve readability, these results are omitted. In this paper, we focus on the frequencies of the responses, which clearly illustrate the interaction path of dual-frequency perturbations on the flame response. Therefore, the corresponding concise expressions are sufficient for us.

The linear solutions can be rewritten as

$$\xi_1(r,t)=\mu_1(r)\cos\left[St_1t+\chi_1(r)\right] \tag{B 12}$$

$$\xi_2(r,t,\delta)=\mu_2(r,\delta)\cos\left[St_2t+\chi_2(r,\delta)\right] \tag{B 13}$$

The self-nonlinear solutions are

$$\begin{aligned}\xi_{1,1}(r,t)&=\xi_{\mathrm{I},1,1}(r,t)+\xi_{\mathrm{II},1,1}(r)\\&=\mu_{\mathrm{I},1,1}(r)\cos\left[2St_1t+\chi_{\mathrm{I},1,1}(r)\right]+\mu_{\mathrm{II},1,1}(r)\end{aligned} \tag{B 14}$$

$$\begin{aligned}\xi_{2,2}(r,t,\delta)&=\xi_{\mathrm{I},2,2}(r,t,\delta)+\xi_{\mathrm{II},2,2}(r)\\&=\mu_{\mathrm{I},2,2}(r,\delta)\cos\left[2St_2t+\chi_{\mathrm{I},2,2}(r,\delta)\right]+\mu_{\mathrm{II},2,2}(r)\end{aligned} \tag{B 15}$$

$$\begin{aligned}\xi_{1,1,1}(r,t)&=\xi_{\mathrm{I},1,1,1}(r,t)+\xi_{\mathrm{II},1,1,1}(r,t)\\&=\mu_{\mathrm{I},1,1,1}(r)\cos\left[St_1t+\chi_{\mathrm{I},1,1,1}(r)\right]\\&\quad+\mu_{\mathrm{II},1,1,1}(r)\cos\left[3St_1t+\chi_{\mathrm{II},1,1,1}(r)\right]\end{aligned} \tag{B 16}$$

$$\begin{aligned}\xi_{2,2,2}(r,t,\delta)&=\xi_{\mathrm{I},2,2,2}(r,t,\delta)+\xi_{\mathrm{II},2,2,2}(r,t,\delta)\\&=\mu_{\mathrm{I},2,2,2}(r,\delta)\cos\left[St_2t+\chi_{\mathrm{I},2,2,2}(r,\delta)\right]\\&\quad+\mu_{\mathrm{II},2,2,2}(r,\delta)\cos\left[3St_2t+\chi_{\mathrm{II},2,2,2}(r,\delta)\right]\end{aligned} \tag{B 17}$$

and the three mutual-nonlinear solutions are

$$\begin{aligned}\xi_{1,2}(r,t,\delta)&=\xi_{\mathrm{I},1,2}(r,t,\delta)+\xi_{\mathrm{II},1,2}(r,t,\delta)\\&=\mu_{\mathrm{I},1,2}(r,\delta)\cos\left[(St_1+St_2)t+\chi_{\mathrm{I},1,2}(r,\delta)\right]\\&\quad+\mu_{\mathrm{II},1,2}(r,\delta)\cos\left[|St_1-St_2|t+\chi_{\mathrm{II},1,2}(r,\delta)\right]\end{aligned} \tag{B 18}$$

$$\begin{aligned}\xi_{1,1,2}(r,t,\delta)&=\xi_{\mathrm{I},1,1,2}(r,t,\delta)+\xi_{\mathrm{II},1,1,2}(r,t,\delta)+\xi_{\mathrm{III},1,1,2}(r,t,\delta)\\&=\mu_{\mathrm{I},1,1,2}(r,\delta)\cos\left[St_2t+\chi_{\mathrm{I},1,1,2}(r,\delta)\right]\\&\quad+\mu_{\mathrm{II},1,1,2}(r,\delta)\cos\left[(2St_1+St_2)t+\chi_{\mathrm{II},1,1,2}(r,\delta)\right]\\&\quad+\mu_{\mathrm{III},1,1,2}(r,\delta)\cos\left[|2St_1-St_2|t+\chi_{\mathrm{III},1,1,2}(r,\delta)\right]\end{aligned} \tag{B 19}$$

$$\begin{aligned}\xi_{1,2,2}(r,t,\delta)&=\xi_{\mathrm{I},1,2,2}(r,t)+\xi_{\mathrm{II},1,2,2}(r,t,\delta)+\xi_{\mathrm{III},1,2,2}(r,t,\delta)\\&=\mu_{\mathrm{I},1,2,2}(r)\cos\left[St_1t+\chi_{\mathrm{I},1,2,2}(r)\right]\\&\quad+\mu_{\mathrm{II},1,2,2}(r,\delta)\cos\left[(2St_2+St_1)t+\chi_{\mathrm{II},1,2,2}(r,\delta)\right]\\&\quad+\mu_{\mathrm{III},1,2,2}(r,\delta)\cos\left[|2St_2-St_1|t+\chi_{\mathrm{III},1,2,2}(r,\delta)\right]\end{aligned} \tag{B 20}$$

where $\mu$ ($\mu\geq 0$, where $\mu$ is a function of $St$) and $\chi$ are functions of spatial location and represent the amplitudes and phases associated with each-order solution terms, respectively. The subscripts "I", "II", and "III" (if present) denote the properties corresponding to the first, second, and third disturbance frequencies included in the solution, respectively.